\documentclass{emulateapj}

\newcommand{\kms}{\mbox{km s$^{-1}$}}

\slugcomment{Draft June 24  2010 }

\shorttitle{Disentangling the circumnuclear environs of Centaurus A. II}
\shortauthors{D. Espada et al.}

\begin{document}

\title{Disentangling the circumnuclear environs of Centaurus A: \\
II. On the nature of the broad absorption line}

\author{
D. Espada \altaffilmark{1,2,3}, A. B. Peck \altaffilmark{4,5,6}, S. Matsushita \altaffilmark{1}, 
K. Sakamoto \altaffilmark{1}, 
C. Henkel \altaffilmark{7}, 
D. Iono \altaffilmark{8}, 
F. P. Israel \altaffilmark{9},
S. Muller \altaffilmark{10},
G. Petitpas\altaffilmark{5}, 
Y. Pihlstr{\"o}m \altaffilmark{11,6}, 
G. B. Taylor \altaffilmark{11,6}, and 
D.~V. Trung \altaffilmark{12,1}
}

\altaffiltext{1}{Academia Sinica, Institute of Astronomy and Astrophysics, P.O. Box 23-141, Taipei 10617, Taiwan. } 
\altaffiltext{2}{Harvard-Smithsonian Center for Astrophysics, 60 Garden St., Cambridge, MA 02138, USA; despada@cfa.harvard.edu.}
\altaffiltext{3}{Instituto de Astrof{\'i}sica de Andaluc{\'i}a - CSIC, Apdo. 3004, 18080 Granada, Spain.}
\altaffiltext{4}{Joint ALMA Office, Av. El Golf 40, Piso 18, Las Condes, Santiago, Chile.}
\altaffiltext{5}{Harvard-Smithsonian Center for Astrophysics, Submillimeter Array, 645 North A`ohoku Place, Hilo, 96720, USA.}
\altaffiltext{6}{National Radio Astronomy Observatory, P.O. Box O, Socorro, NM 87801, USA.}
\altaffiltext{7}{Max-Planck-Institut f{\"u}r Radioastronomie, Auf dem H{\"u}gel 69, 53121 Bonn, Germany.} 
\altaffiltext{8}{Nobeyama Radio Observatory, NAOJ, Minamimaki, Minamisaku, Nagano, 384-1305, Japan.}
\altaffiltext{9}{Sterrewacht Leiden, Leiden University, Postbus 9513, 2300 RA Leiden, the Netherlands.}
\altaffiltext{10}{Onsala Space Observatory, SE-43992 Onsala, Sweden.}
\altaffiltext{11}{Department of Physics and Astronomy, MSC07 4220, University of New Mexico, Albuquerque, NM 87131, USA.}
\altaffiltext{12}{Center for Quantum Electronics, Institute of Physics, Vietnam Academy of Science and
Technology, 10 DaoTan, ThuLe, BaDinh, Hanoi, Vietnam.}

\begin{abstract}

We report on atomic gas (\ion{H}{1}) and molecular gas (as traced by CO(2--1)) redshifted absorption features toward the nuclear regions of the closest  powerful radio galaxy, Centaurus A (NGC 5128).
Our \ion{H}{1} observations using the Very Long Baseline Array allow us to discern with unprecedented sub-parsec resolution \ion{H}{1} absorption profiles toward different positions along the 21~cm continuum jet emission in the inner $0\farcs3$ (or 5.4~pc). 
In addition, our CO(2--1) data obtained with the Submillimeter Array probe the bulk of the absorbing molecular gas with little contamination by emission, not possible with previous CO single-dish observations. 
We shed light with these data on the physical properties of the gas in the line of sight, emphasizing the still open debate about the nature of the gas that produces the broad absorption line ($\sim$ 55~\kms). First, the broad \ion{H}{1} line is more prominent toward the central and brightest 21~cm continuum component than toward a region along the jet at a distance $\sim$ 20 mas (or 0.4~pc) further from it. This suggests that the broad absorption line arises from gas located close to the nucleus, rather than from diffuse and more distant gas. 
Second, the different velocity components detected in the CO(2--1) absorption spectrum match well other molecular lines, such as those of HCO$^+$(1--0), except the broad absorption line that is detected in HCO$^+$(1--0) (and most likely related to that of the \ion{H}{1}).
 Dissociation of molecular hydrogen due to the AGN seems to be efficient at distances $r$ $\lesssim$ 10~pc, which might contribute to the depth of the broad \ion{H}{1} and molecular lines. 
\end{abstract}

\keywords{galaxies: elliptical and lenticulars --- galaxies: individual (NGC~5128) --- galaxies: structure --- galaxies: ISM}

\section{Introduction}
\label{introduction}

\setcounter{footnote}{0}

The analysis of absorption lines arising from atomic and molecular gas in galaxies is a powerful technique to study the physical and chemical properties of the interstellar medium (ISM).
Only a handful of nearby galaxies are known to show atomic (\ion{H}{1}) and molecular absorption features. \ion{H}{1}\ and molecular absorption line studies are not only relevant to deduce the properties of the ISM along specific lines of sight in nearby galaxies \citep[e.g.][]{2007ApJ...654L.111G}, but are also essential for an interpretation of similar data toward higher redshift  galaxies \citep[e.g.][]{2006A&A...458..417M,2008Ap&SS.313..321C,2009A&A...500..725H}.
  Studying both \ion{H}{1} and molecular gas absorption lines at the same time can reveal the interface between both components of the ISM at the circumnuclear regions where dissociation of molecular gas is expected as a result of energetic processes triggered by Active Galactic Nuclei (AGNs).

At a distance of $D$ $\simeq$ 3.8~Mpc \footnote{This distance is used throughout this paper.} (\citealt{2009arXiv0911.3180H}), Centaurus~A (Cen~A, NGC\,5128) is the nearest giant elliptical exhibiting nuclear activity, with powerful radio jets extending almost 10$\arcdeg$ on the sky and with prominent atomic and molecular emission as well as absorption features detected toward its nuclear regions \citep{1998A&ARv...8..237I}. It is thus a unique source to study the molecular to atomic interface. At this distance, high spatial resolution can be achieved: 1\arcsec\ corresponds to 18~pc projected linear size. For a comprehensive general review of Cen~A, see \citet{1998A&ARv...8..237I}.

One of the most prominent features in Cen~A is the dust lane along its minor axis, which is composed of a large amount of gas and dust within a warped disk-like structure seen nearly edge-on. An \ion{H}{1} mass of (10 $\pm$ 3) $\times$ 10$^8$ M$_\odot$ extends out to a radius of at least $r$ = 7\arcmin\ (or about 7~kpc), while it is observed to be mostly absent within the inner arcmin (or 1~kpc projected linear size) \citep{1990AJ.....99.1781V,1994ApJ...423L.101S,1998A&ARv...8..237I}. A comparable amount of molecular gas is contained within the inner $r$ $\simeq$ 2\arcmin\ (2~kpc), M$\rm _{H_2}$ $\simeq$ 4.2 $\times$10$^8$~M$_\odot$ \citep{1990ApJ...363..451E}. About 30\% -- 40\% of the total molecular gas content is located within the inner 1\arcmin\ (1~kpc), M$\rm _{H_2}$ = (8.1 $\pm$ 0.4) $\times$10$^7$~M$_\odot$  \citep[][hereafter Paper I]{2009ApJ...695..116E}. 
A circumnuclear disk of molecular gas is resolved in the inner kiloparsec, with a size of 24\arcsec $\times$ 12\arcsec\ (430 $\times$ 215 pc) in diameter, and with different kinematics with respect to the molecular gas at larger radii (Paper I).

Along the NE -- SW direction (perpendicular to the circumnuclear molecular gas disk), there exists a spectacular radio-continuum jet extending from the nucleus (pc scale) to the outer radio lobes (hundreds of kpc scale). Subparsec-scale VLBI (Very Large Baseline Interferometry, including the VSOP - VLBI space observatory programme) high resolution radio-continuum observations (e.g. at 2.3, 4.8, 8.4, 22.2 and 43~GHz) show that the structure of Cen A is rather complex, consisting of a bright nuclear jet and a fainter counterjet, as well as a bright central component  \citep[e.g.][]{1998AJ....115..960T,2001ApJ...546..210T,2006PASJ...58..211H}. Note that this component is not the nucleus itself since the nucleus is not bright at frequencies below 10 -- 20~GHz due to synchrotron self-absorption and free-free absorption from a disk or torus of ionized gas \citep{1996ApJ...466L..63J,2001ApJ...546..210T}. The compact core becomes visible at mm/submm wavelengths \citep{1997ApJ...475L..93K} where its spectral index is quite flat \citep[see Fig.~6 in ][]{2007A&A...471..453M}.

\ion{H}{1} absorption lines in this source were found for the first time four decades ago (e.g. \citealt{1970ApJ...161L...9R}). Since then, VLA aperture synthesis observations with resolutions of the order of $\sim$ 10\arcsec\ \citep{1983ApJ...264L..37V,2002ApJ...564..696S} have shown \ion{H}{1} absorption lines in different regions along the jet. 
 They are distributed toward both the core and the inner jet (extended along 30\arcsec ) and are composed of a set of lines with widths of typically 10 \kms . The most prominent \ion{H}{1} absorption component is located close to the systemic velocity of the galaxy ($V_{\rm LSR}$ = 541.6
 $\pm$ 2~\kms ,  \citealt{1998A&ARv...8..237I})\footnote{All velocities in this paper are radio LSR.
Conversion from LSR to heliocentric velocity is  $V_{\rm Hel}$ = $V_{\rm LSR}$ + 2.4 \kms .}. 
 Toward the compact bright continuum component, two redshifted lines are seen at $V$ = 578~\kms\  and 598~\kms, as well as a broad absorption component (extending up to $V$ = 620~\kms\ with a width of $\Delta V$ $\simeq$ 55~\kms). \citet{2002ApJ...564..696S} showed that the latter is only present toward the central component, but not along the other regions of the jet. This sets an upper limit to the size of the absorbing material of about 20\arcsec, or 360~pc as seen in projection. 

Molecular absorption lines  from cm to submm wavelengths have also been detected for several species. 
At cm wavelengths narrow molecular lines (a few \kms\ width) close to the systemic velocity have been observed in several molecules such as OH,  H$_2$CO, C$_3$H$_2$ and NH$_3$ \citep{1976PASA...3...63G,1976MNRAS.175P...9G,1990ApJ...364...94S,1995ApJ...448L.123V,2005Ap&SS.295..249V}. 
Single-dish observations have been performed at mm/submm wavelengths for the lowest transitions of CO, HCN, HNC, HCO$^+$, H$^{13}$CO$^+$, N$_2$H$^+$, C$_2$H, CN and CS \citep{1990ApJ...365..522E,1990A&A...227..342I,1997A&A...324...51W,1999ApJ...516..769E}.
In addition to narrow lines close to the systemic velocity, both narrow and broad absorption features are seen at redshifted velocities in HCO$^+$(1--0), OH 18cm, and to a minor extent in HCN(1--0), HNC(1--0) and CS(2--1), covering a velocity range from $\sim$ 570 to 620 \kms\ \citep[e.g.][]{1990ApJ...364...94S,1990ApJ...363..451E,1991A&A...245L..13I,1997A&A...324...51W}. 

Although there is a consensus that narrow absorption lines with velocities close to the systemic velocity represent gas far from the nucleus, a dichotomy exists regarding the nature of the broad redshifted absorption line. It has been suggested that it might be: i) warm and dense gas close to the nucleus and probably falling into it \citep[e.g.][]{1983ApJ...264L..37V,1991A&A...245L..13I}, and ii) cold diffuse gas above and below a warped disk \citep{1999ApJ...516..769E}.
In this paper we aim to shed light on the model that better represents the different gaseous components (both atomic and molecular) producing such absorption lines, and their physical properties as a result of their location with respect to the nucleus. In particular, we emphasize the origin of the broad line component.

We have carried out \ion{H}{1} observations using the Very Large Baseline Array (VLBA) and the phased Very Large Array (VLA)\footnote{The VLBA and VLA are operated by the National Radio Astronomy Observatory, a facility of the National Science Foundation operated under cooperative agreement by Associated Universities, Inc.} which allow us to derive the properties of the main \ion{H}{1} absorption features seen toward the continuum source at 21~cm with unprecedented sub-parsec resolution (versus the $\sim$ 180~pc resolution in past VLA experiments). 
This is complemented by CO(2--1) observations (and isotopologues) at the Submillimeter Array (SMA\footnote{The Submillimeter Array is a joint project between the Smithsonian Astrophysical Observatory and the Academia Sinica Institute of Astronomy and Astrophysics, and is funded by the Smithsonian Institution and the Academia Sinica.}, \citealt{2004ApJ...616L...1H}) (Paper I) from which we can extract a CO(2--1) absorption profile with minimal contamination by emission. 
We introduce our VLBA \ion{H}{1} and SMA CO(2--1)  observations and the data reduction in \S~\ref{observationReduction}, where we present the continuum maps and absorption spectra. 
In \S~\ref{resultsHI} we focus on the identification of the individual \ion{H}{1} absorption components and the physical properties of their corresponding regions. We present our results on the CO(2--1) and $^{13}$CO(2--1) absorption lines in \S~\ref{resultsCO} and compare the \ion{H}{1} absorption lines with those of molecular lines in \S~\ref{subsec:comparison}. We discuss the results in \S~\ref{discussion} and summarize our findings in \S~\ref{conclusion}.

\section{Observations and Data Reduction}
\label{observationReduction}

The array configuration, phase center, primary and synthesized beam, achieved noise and spectral resolution of our  VLBA \ion{H}{1} and SMA CO(2--1) observations are shown in Table~\ref{tbl-2}.
In the following we explain in more detail our observational strategies and data reduction.

\subsection{VLBA \ion{H}{1} Observations}
\label{VLBA-HI}

The VLBA \ion{H}{1}\ observations were carried out on 26 November 1994.  Due to its low declination (Dec.~$\simeq -43\arcdeg$), Cen\,A was observable for 2 hours with 8 of the 10 VLBA antennas and the phased VLA.  The observations were made using 2 IFs of 2 MHz each.  The IFs were centered on the redshifted  \ion{H}{1}\ line ($\nu_{\rm rest}$ = 1420.405752 MHz, at a velocity $V_{\rm LSR}$ = 541.6 km s$^{-1}$). 
Global fringe-fitting was performed using AIPS\footnote{http://www.aips.nrao.edu/cook.html.} task FRING with a
solution interval of 6 minutes. 
Amplitude calibration was derived
using the antenna gain and system temperature recorded
at each antenna. Bandpass calibration was done using
the calibrator 3C273 (J1229+0203). 
The continuum was obtained from line-free channels.
Continuum subtraction was done in \verb!Difmap! 
\citep{1995S} using a model of \verb!CLEAN! components made from several line-free channels.  Subsequent editing and imaging of all data was also done using \verb!Difmap!, using natural weighting. The resulting Half Power Beam Width (HPBW) is 36.3 $\times$ 7.2~mas (0.65 $\times$ 0.13~pc). 
The velocity resolution in these observations is 3.3~\kms, and the total velocity coverage of a single IF is $\sim$ 430~\kms. 

\subsection{SMA CO(2--1) Observations} 
\label{CO-SMA}

Details of the CO(2--1) observations using the SMA in its compact configuration are reported in Paper I.  CO(2--1), $^{13}$CO(2--1) and C$^{18}$O(2--1) lines were simultaneously observed.
We used two independent sources, 3C273 and Callisto, to check the consistency of our bandpass calibration. Both solutions were nearly identical in most of the spectral windows, and in particular where the CO(2--1) lines were located. The continuum toward Cen\,A at 1.3 mm was found to consist of an unresolved source with a flux of $S\rm _{\rm 1.3mm}$ = 5.9 $\pm$ 1.0 Jy beam$^{-1}$ at the AGN position which was used to calibrate the data of Cen\,A itself (Paper I). 
No continuum subtraction was performed. 
The velocity resolution  is  1 \kms.

In order to minimize undesired emission, we obtain the CO(2--1) absorption spectrum by directly fitting the amplitude of the interferometric
visibilities using a point source model fixed at the position of the unresolved continuum source. This was done with the task \verb!UV_FIT! in the \verb!GILDAS!\footnote{http://www.iram.fr/IRAMFR/GILDAS.} reduction package. 
The CO(2--1) spectrum extracted using visibilities from all baselines still shows some emission at velocities $V$ = 420 -- 510~\kms\ and  $V$ = 620 -- 680~\kms . This corresponds to extended emission  from the circumnuclear regions (Paper I).
A new fit to the visibility data with projected baseline lengths $\ge$ 50~m yielded a flatter spectral baseline. This is due to a combination of the spatial filtering capabilities of the interferometer, beam smearing and probably a lack of CO(2--1) emission toward the central $\sim$ 5\arcsec\ ($\sim$ 90~pc). The typical extent that we take into account by removing the shorter baselines can be roughly estimated as $\le \lambda / D$ $\simeq$ 4\arcsec\ ($\sim$ 72~pc), where
$\lambda$ is the wavelength of the observations and $D$ the average baseline length.  

\begin{figure}
\centering
\includegraphics[width=10cm]{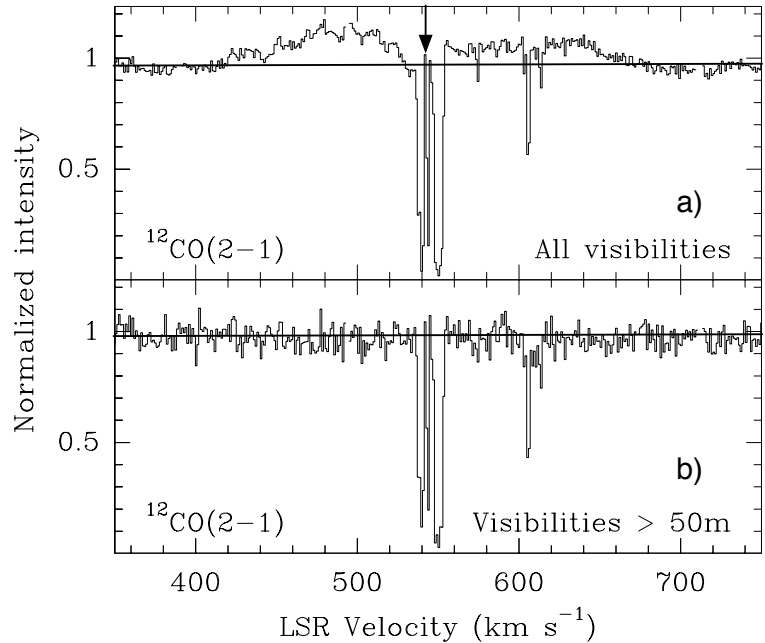}
\caption{ $a)$ CO(2--1) normalized intensity spectra obtained from all the visibilities using a fit in the UV plane assuming a point-like feature for the continuum emission. It clearly fails to eliminate most of the emission arising from the circumnuclear molecular gas \citep{2009ApJ...695..116E}. The arrow indicates the systemic velocity of the galaxy ($V$ = 541.6~\kms ).
$b)$ Same as $a)$ but only using visibilities with projected baselines larger than 50 m. The contamination by emission is efficiently diminished. 
\label{fig-spectrumComposition2-1}}
\end{figure}

 We show in Figure~\ref{fig-spectrumComposition2-1} the resulting CO(2--1) spectra with and without visibilities with baseline lengths larger than 50~m. 
For the $^{13}$CO(2--1) and C$^{18}$O(2--1) transitions, however, we retained all visibilities  since there is no significant emission down to our noise level. Their resulting noise per channel, $\sigma =0.1$~Jy, is thus smaller than in the long-baseline only CO(2--1) spectrum ($\sigma = $ 0.3~Jy).

\section{  \ion{H}{1} Absorption Features}
\label{resultsHI}

\subsection{21~cm Continuum: Nuclear Jet}
\label{21cmcontinuum}

\begin{figure}
\centering
\includegraphics[width=10cm,angle=-90]{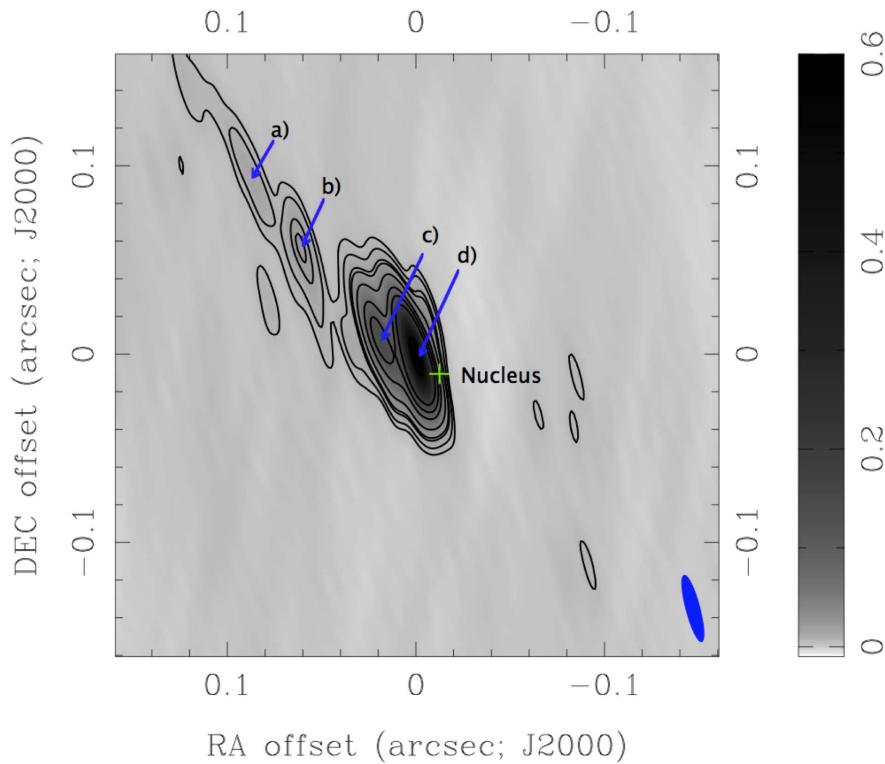}
\caption{Nuclear jet of the  21~cm continuum emission of Cen~A in the inner 0\farcs3 (or 5.4~pc), as observed with the VLBA.  The brightest component has a flux density peak of $S_{\rm 21cm}$ = 600 $\pm$ 30~mJy~beam$^{-1}$. Note that this is not associated with the nucleus itself but with the base of the nuclear jet (\S~\ref{introduction}). The cross indicates the approximate location of the 8.4~GHz continuum core \citep{2001ApJ...546..210T}, assuming that the 1.4~GHz (or 21~cm) brightest component coincides with that of the 2.2~GHz emission. Contours are drawn
at 3, 5, 9, 11, 20, 40, 60, 90 and 250 $\times$ $\sigma$, where $\sigma$ = 2 mJy beam$^{-1}$. The grey scale ranges from $-10$ to 603 mJy beam$^{-1}$. The restoring beam is characterized by a half power beam width HPBW = 36.3 $\times$ 7.2~mas and a position angle P.A. = 14\fdg8, as is shown by the filled ellipse in the lower right of the plot. Note that 1~mas corresponds to 0.018~pc. }
\label{fig5.4}
\end{figure}

The 21~cm continuum map in the inner $0\farcs3$ (5.4~pc) is presented in Figure~\ref{fig5.4} (center).
This high resolution map resolves the bright 21~cm central component ($\sim$ 10\arcsec , or 180 pc) found by \citet{1983ApJ...264L..37V} and \citet{2002ApJ...564..696S} (labeled as ``Nucleus''). We find that this central component further comprises a more compact bright component and a nuclear jet extending up to 0\farcs23 to the NE, with a similar P.A. $\simeq$ 51\arcdeg\ as the inner 21~cm jet extending up to 55\arcsec\ \citep{1983ApJ...264L..37V,2002ApJ...564..696S}. Note that \citet{2005Ap&SS.295..249V}  present a VLBI 18~cm continuum map with a sub-arcsecond resolution where the central bright component is seen. However, the rest of the nuclear jet is not as clearly detected as in our map.

While the peak flux density is 603 mJy~beam$^{-1}$ and the noise level of our maps is 2 mJy beam$^{-1}$, note that the uncertainty of the absolute flux density measurement is larger. Uncertainties in flux calibration are estimated to be about 5\% ($S\rm _{\rm 21cm}$ = 600 $\pm$  30~mJy~beam$^{-1}$). In addition, on-source errors due to deconvolution are expected.  For instance, \citet{2001ApJ...546..210T} estimated that  the corresponding flux density error in their 2.2~GHz VLBA data of Cen~A due to this effect is about 30\% for the brightest components. This source of error is expected to be larger for our lower observing frequency, since the $uv$ coverage is less uniform.

As mentioned in \S~\ref{introduction}, the core of the radio source has a very strongly inverted spectrum since the continuum emission is seen through a circumnuclear ionized gas disk which is opaque at wavelengths longer than 13~cm (or frequencies lower than 2.3~GHz) due to free-free absorption and to self-absorbed synchrotron emission \citep{1996ApJ...466L..63J,2001ApJ...546..210T}.
The compact bright component in Figure~\ref{fig5.4} is thus part of the approaching jet and is offset from the actual AGN. 
 \citet{2001ApJ...546..210T} showed that there is an absolute offset of
12.5~mas to the E and 10.5~mas to the N between the  8.4~GHz and 2.2~GHz central components. This corresponds to a separation of 0.3~pc along the jet axis. 
We show in Fig.~\ref{fig5.4} the estimated location of the AGN (8.4~GHz central component), assuming that the bright component of our 1.4~GHz map coincides with that of the 2.2~GHz map, and also that position variability between the different observing dates is not significant.
We do not detect any continuum emission from the counter-jet toward the SW to the 2~mJy beam$^{-1}$ noise level of our VLBA maps. 

\subsection{Identification of \ion{H}{1} Absorption Features}
\label{sub:HIabsorptionfeatures}

\begin{figure}
\centering
\includegraphics[width=7cm,angle=-90]{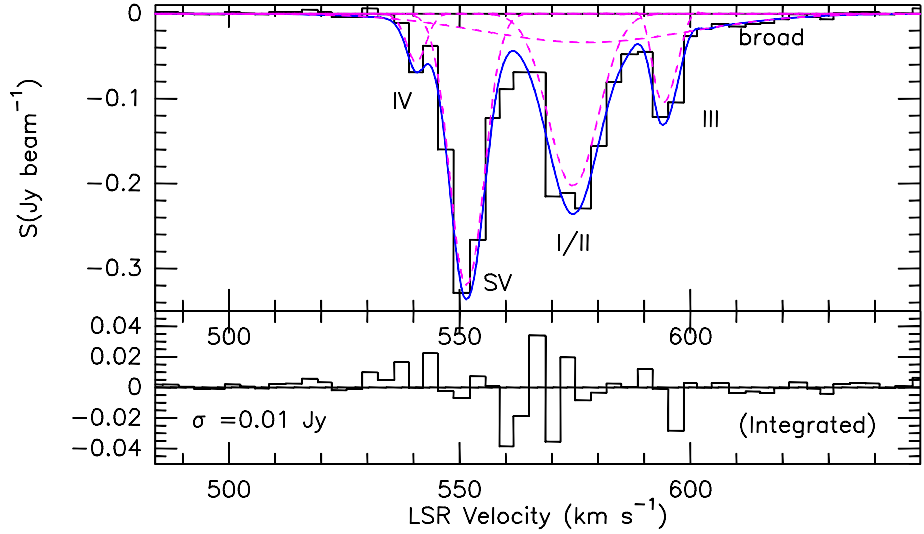}

\caption{ \ion{H}{1} absorption profile of Cen~A, integrated over the entire 21~cm continuum region probed by our VLBA experiment (Fig.~\ref{fig5.4}).  The profile is mainly composed of five lines: IV, SV, I/II and III, plus an underlying broader line with a width of 53 $\pm$ 10~km~s$^{-1}$, whose wing is clearly seen at $V$ $>$ 600~\kms. The velocity resolution is 3.3 \kms\ and the rms noise of 3 mJy beam$^{-1}$ per channel. 
The solid line is the sum of the Gaussian fits to the individual lines (dashed lines, as specified in Table~\ref{tbl-5}). The plot at the bottom shows the residual from this fit, with a rms $\sigma$ = 0.01~Jy.
\label{fig6-2} }
\end{figure}

Figure \ref{fig6-2} shows the \ion{H}{1} spectrum integrated over the entire continuum source. We find 5 main \ion{H}{1} absorption lines: four relatively narrow lines and an underlying broad line.
These absorption lines, all red-shifted  with respect to the systemic velocity, were detected using the VLA by \citet{1983ApJ...264L..37V} and \citet{2002ApJ...564..696S}, although toward an unresolved component of size $\sim$ 8\arcsec\ -- 10\arcsec . We adopt the nomenclature used in \citet{2002ApJ...564..696S} for the most prominent lines (in roman numerals, IV, SV, I, II and III).

 We derive from Gaussian fits the mean velocities ($V$), peak flux densities ($S$) and Full Widths at Half Maximum (FWHM) of  the different lines in our \ion{H}{1} profile, as shown in Table~\ref{tbl-5}. 
We first fit the broad line using the redshifted wing at $V$ $>$ 600~\kms , and then fit the remaining lines individually. An exception is line I/II, which is a blend of lines I and II  in the \ion{H}{1} spectra of  \citet{2002ApJ...564..696S}. Being almost unresolved with our 3.3~\kms\ resolution, we decided to compute them together.
The \ion{H}{1} absorption features in our spectra are located at $V$ = 541, 552, 575 and 595 ($\pm$ 1) \kms\  (IV, SV, I/II, III, as outlined in Figure~\ref{fig6-2}). The most prominent lines show FHWM values of about 5 -- 10~\kms . A fit to the broad absorption line yields a FWHM $=$ 53 $\pm$ 10~\kms\ centered at 578 $\pm$ 4~\kms . The final fit and the resulting residual from the observed \ion{H}{1} profile are presented in  Figure~\ref{fig6-2}.

The great advantage of our VLBA data is that we can obtain the \ion{H}{1} absorption spectrum toward different positions over the continuum emission. Figure~\ref{fig5.4}  shows the four most prominent continuum sources where we obtained \ion{H}{1} absorption profiles. Their relative offsets with respect to the phase center are: ($\Delta\alpha$,$\Delta\delta$) = (+73,+69)~mas, (+63,+58)~mas and (+20,+10)~mas  (namely, positions $a$, $b$ and $c$), corresponding to separations of 1.8~pc, 1.5~pc and 0.4~pc respectively; as well as toward the brightest component at (+1,$-1$)~mas (namely, position $d$). The continuum flux density peaks of the different components are $S\rm _{\rm 21cm}$ = 11, 23, 145 and 603~mJy~beam$^{-1}$, respectively. 

\begin{figure}
\centering
\includegraphics[width=5cm,angle=-90]{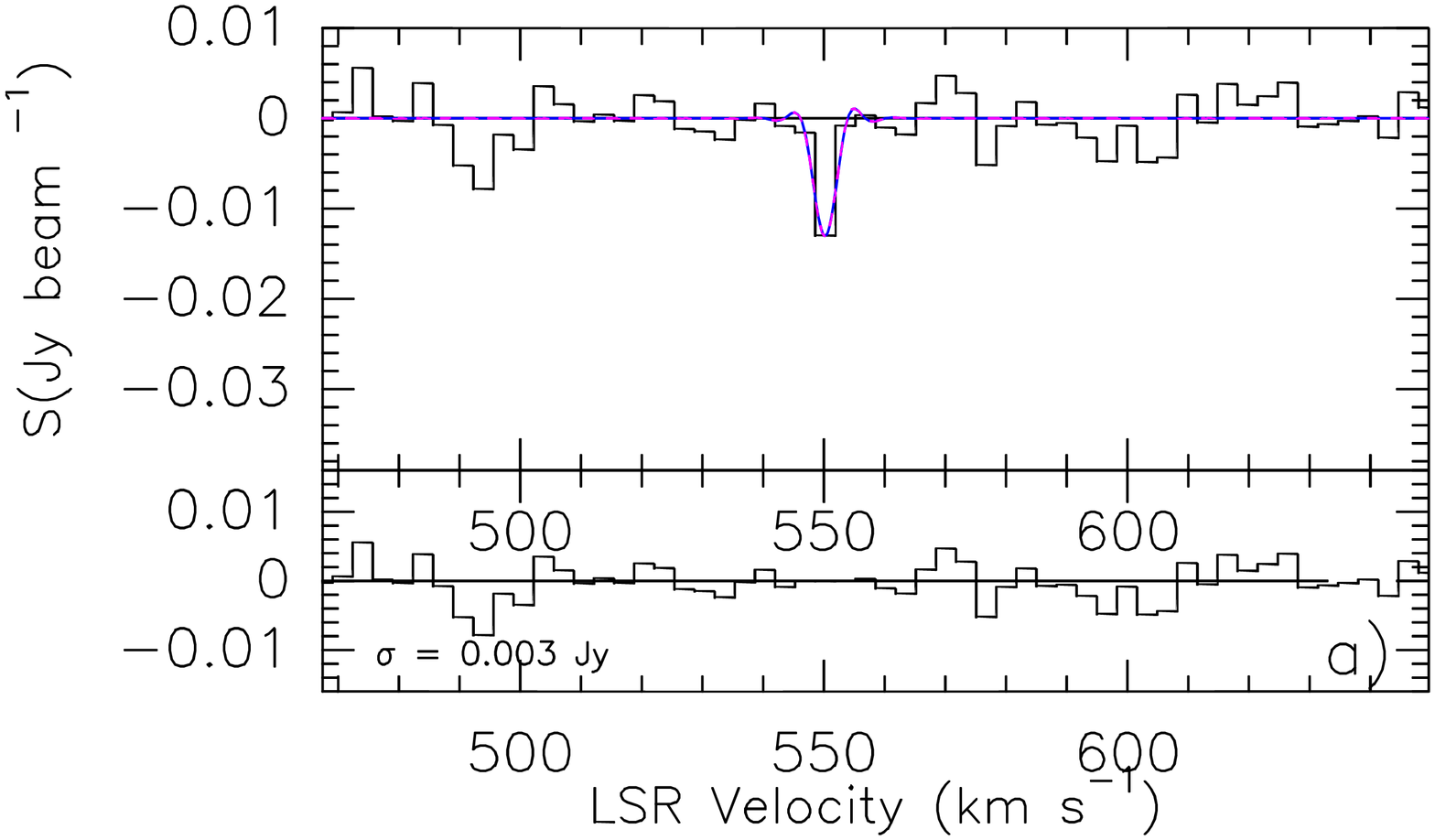}
\includegraphics[width=5cm,angle=-90]{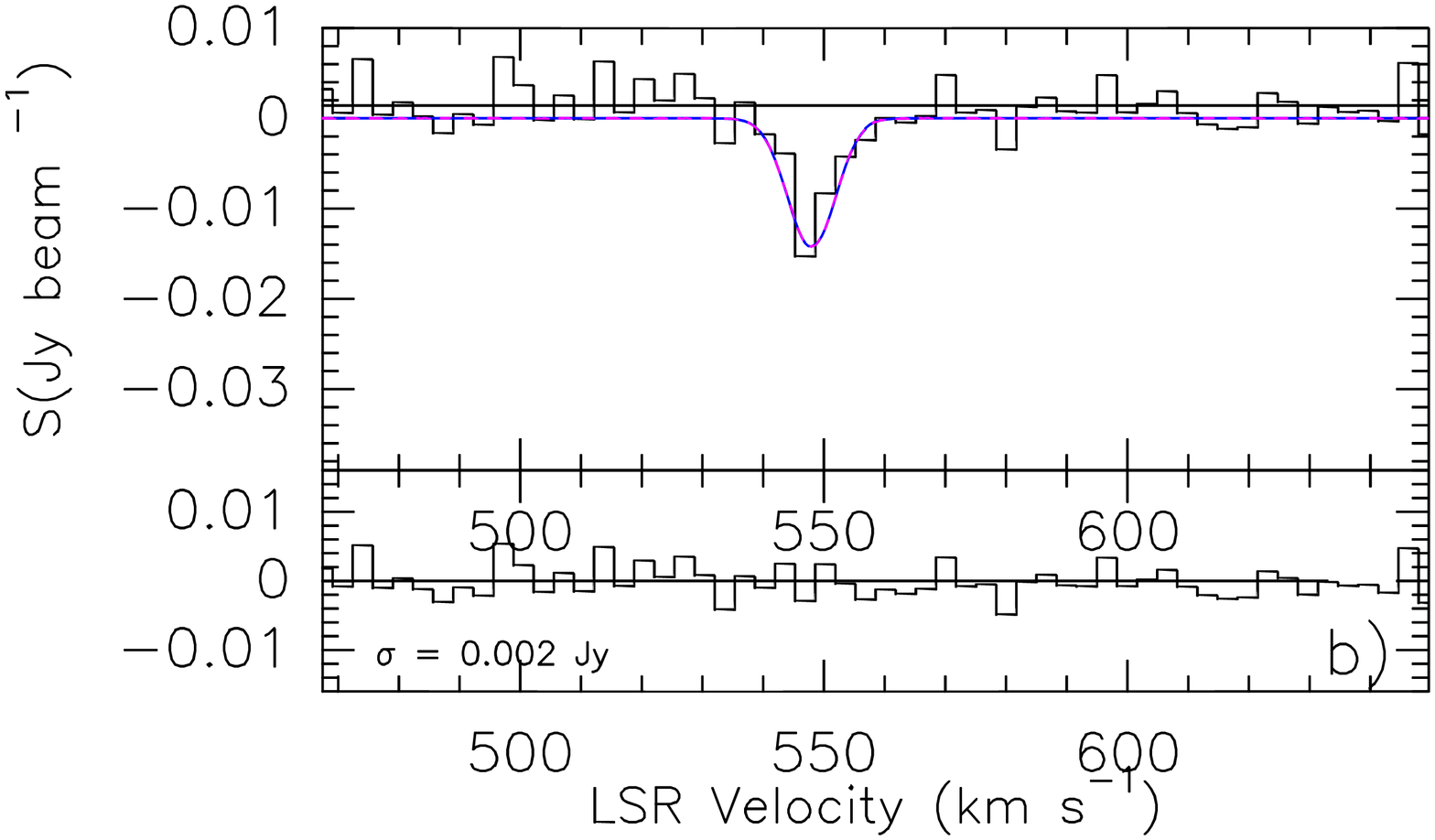}
\includegraphics[width=5cm,angle=-90]{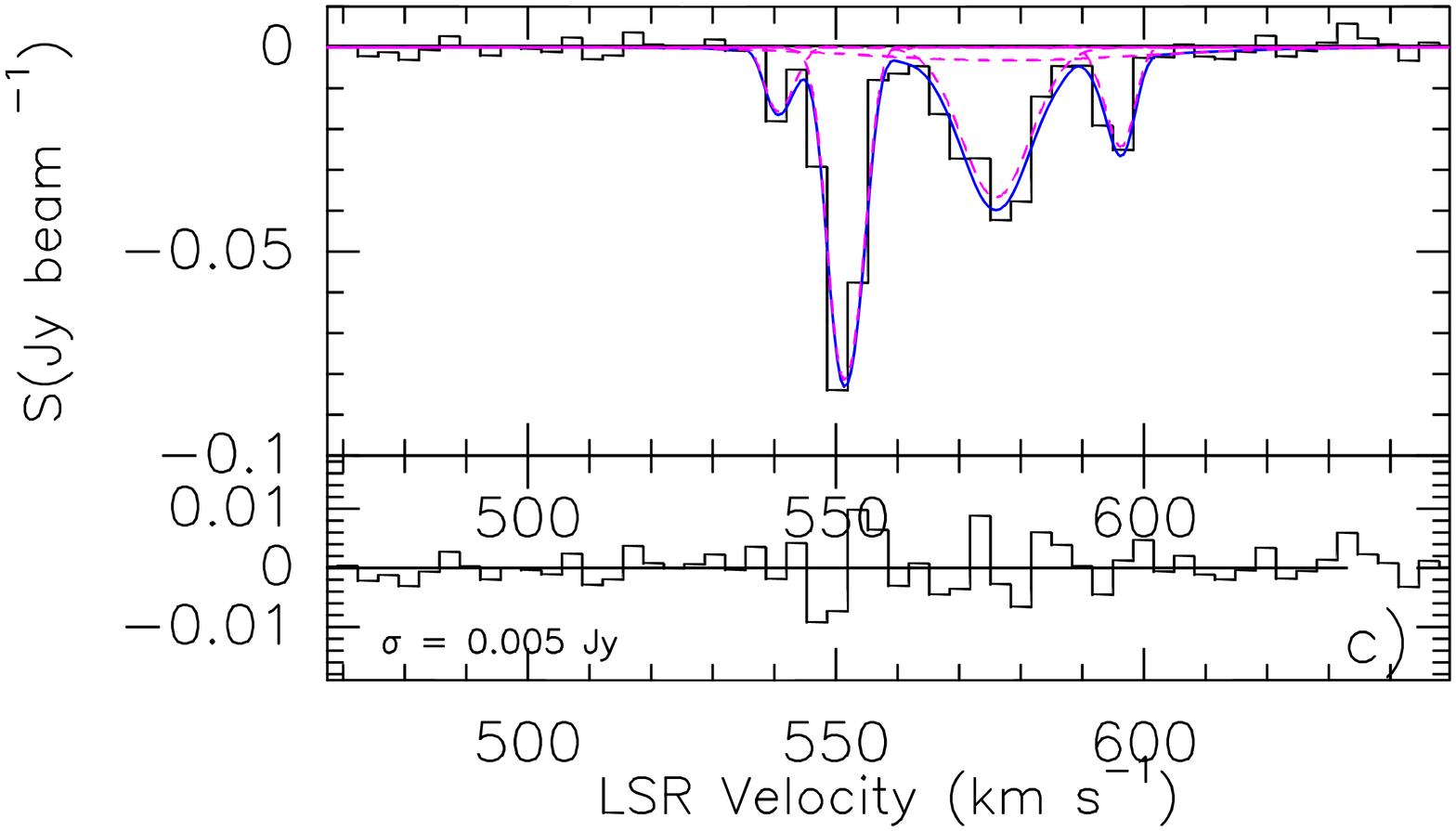}
\includegraphics[width=5cm,angle=-90]{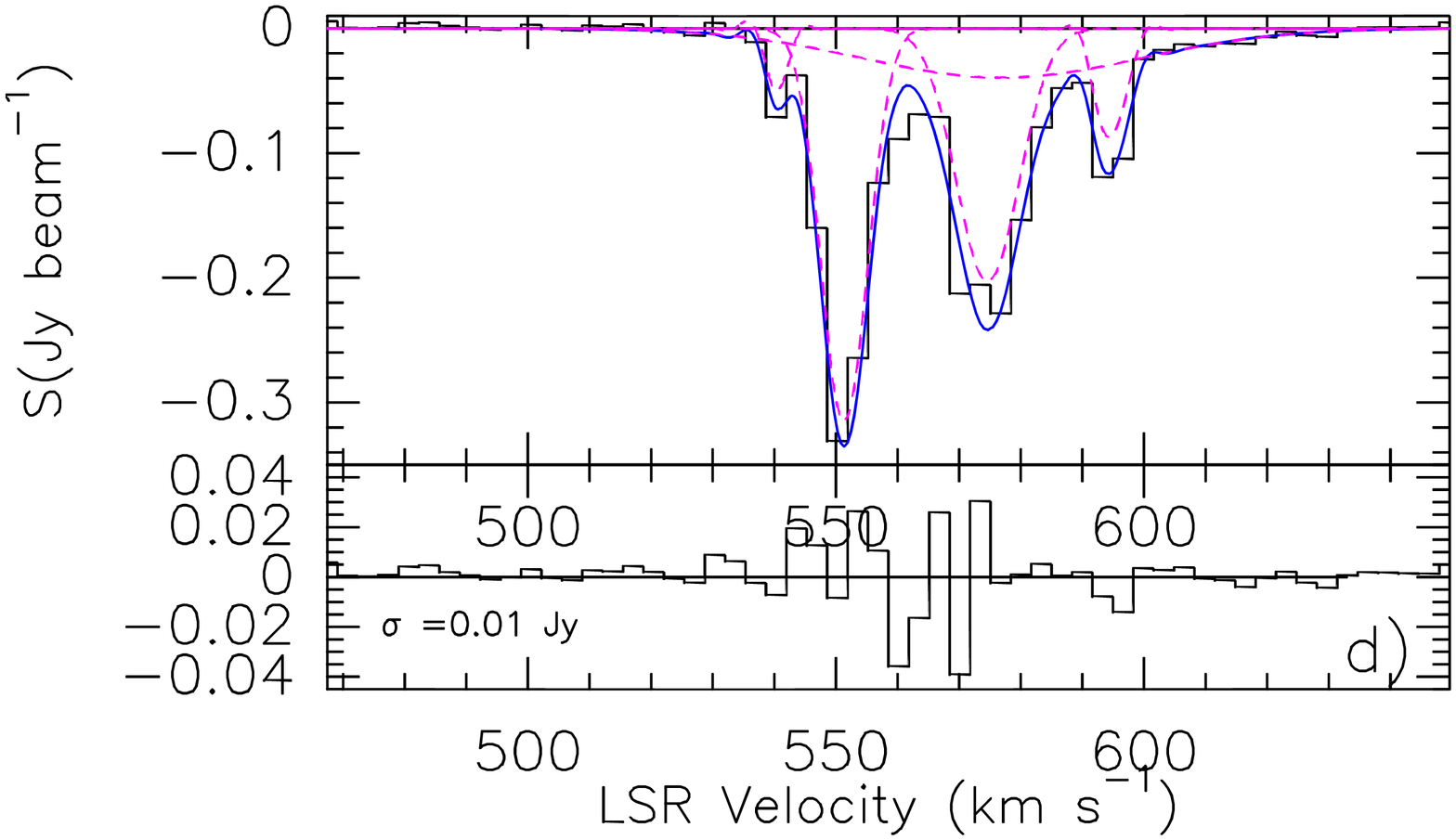}

\caption{ \ion{H}{1} absorption lines toward single pixels at positions $a$, $b$, $c$ and $d$ as outlined in Figure~\ref{fig5.4} (see bottom right of each panel), as well as the individual Gaussian fits and underlying final fitted curve.   The residuals from the fit are characterized by a $\sigma$ as indicated on the lower left of each panel. 
\label{fig6-6}}
\end{figure}

While we detect \ion{H}{1} absorption features only close to the systemic velocity (components IV and SV) in positions $a$ and $b$, in position $c$ and $d$ the profiles show two additional redshifted \ion{H}{1} absorption components I/II and III at $V$ = 574 -- 576~\kms\ and 595 -- 596~\kms\ (Figure~\ref{fig6-6}).  An additional shallow redshifted wing in position $d$ is clearly seen at velocities $V$ $>$ 600~\kms , which corresponds to  the broad absorption line. 
The fitted Gaussian parameters are presented in Table~\ref{tbl-6}, and the fits are shown in Figure~\ref{fig6-6}. 
To fit these lines, we set as initial values the velocities  and FWHM found for the \ion{H}{1} absorption lines integrated over the entire continuum. 
Since the flux densities refer to their corresponding continuum emission, we calculated next the optical depths and \ion{H}{1} column densities in order to compare between different positions.

\subsection{\ion{H}{1}\ Optical Depths and Column Densities }
\label{sub:HIcoldens}

We calculated the   \ion{H}{1} optical depths, $\tau_{\rm HI}$, as
$
I_{\rm obs}[Jy] = I_{\rm cont} e^{-\tau_{\rm HI}},
$
where $I _{\rm obs}$ is the observed flux in absorption and $I _{\rm cont}$ is the measured continuum flux. Absolute flux uncertainties do not have any effect on the optical depth since it is derived from the $I _{\rm obs}$/$I _{\rm cont}$ ratio.
In columns 5 and 6 of Table~\ref{tbl-5} and ~\ref{tbl-6} we present the \ion{H}{1} optical depth peak (peak $\tau_{\rm HI} $) and integrated optical depths ($\int \tau_{\rm HI} dV$) integrated over the entire continuum source and for each independent position, respectively.
The upper limits corresponding to non-detected lines are calculated as 3 $\sigma$ ($\Delta V$ $\delta v$)$^{1/2}$,
where $\Delta V$ is the FWHM in Table~\ref{tbl-5} and $\delta v$ is the channel spacing of 3.3~\kms. 
 
\begin{figure}
\centering
\includegraphics[width=9cm,angle=-90]{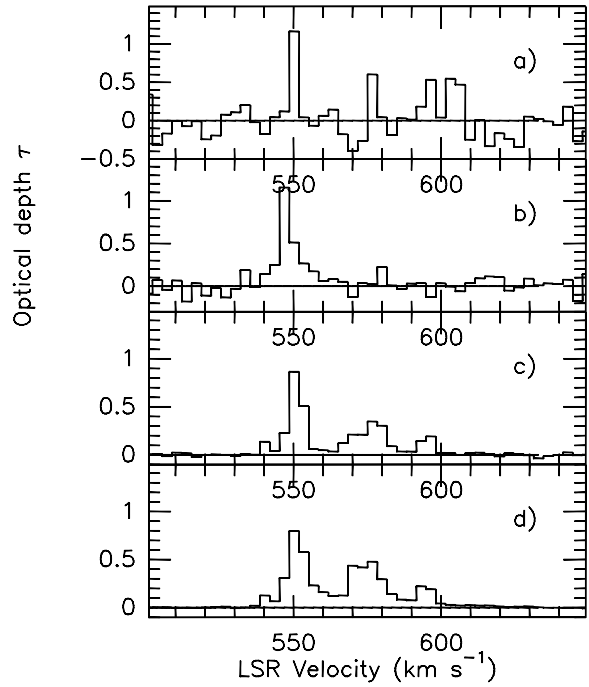}

\caption{  \ion{H}{1}\ optical depths for the four different positions along the 21~cm jet ($a$, $b$, $c$) and brightest component ($d$),
as outlined in Figure~\ref{fig5.4}. Positions are indicated at the upper right of each panel.  \label{fig6-4}}
\end{figure}

 In Figure~\ref{fig6-4} we plot $\tau_{\rm HI}$ toward the four different positions $a$, $b$, $c$ and $d$ along the 21~cm nuclear jet. 
The main absorption line is component SV, with peak $\tau_{\rm HI}$ $\geq$ 0.75 seen at all positions. No trend is seen as a function of distance from the nucleus. On the other hand, the red-shifted velocity component I/II seems to be preferentially deeper if closer to the brightest component (position $d$).  
Component I/II is characterized by peak $\tau_{\rm HI}$ $=$ 0.29 $\pm$ 0.03 in position $c$, while peak $\tau_{\rm HI}$ $=$ 0.41 $\pm$ 0.02 in position $d$. 
The $3\sigma$ upper limit obtained in position $b$ is peak $\tau_{\rm HI}$ $<$ 0.3. However, we would have expected at least a tentative detection for component I/II in position $b$ if of comparable depth as in position $c$ or $d$. 

\begin{figure}
\centering
\includegraphics[width=6.5cm,angle=-90]{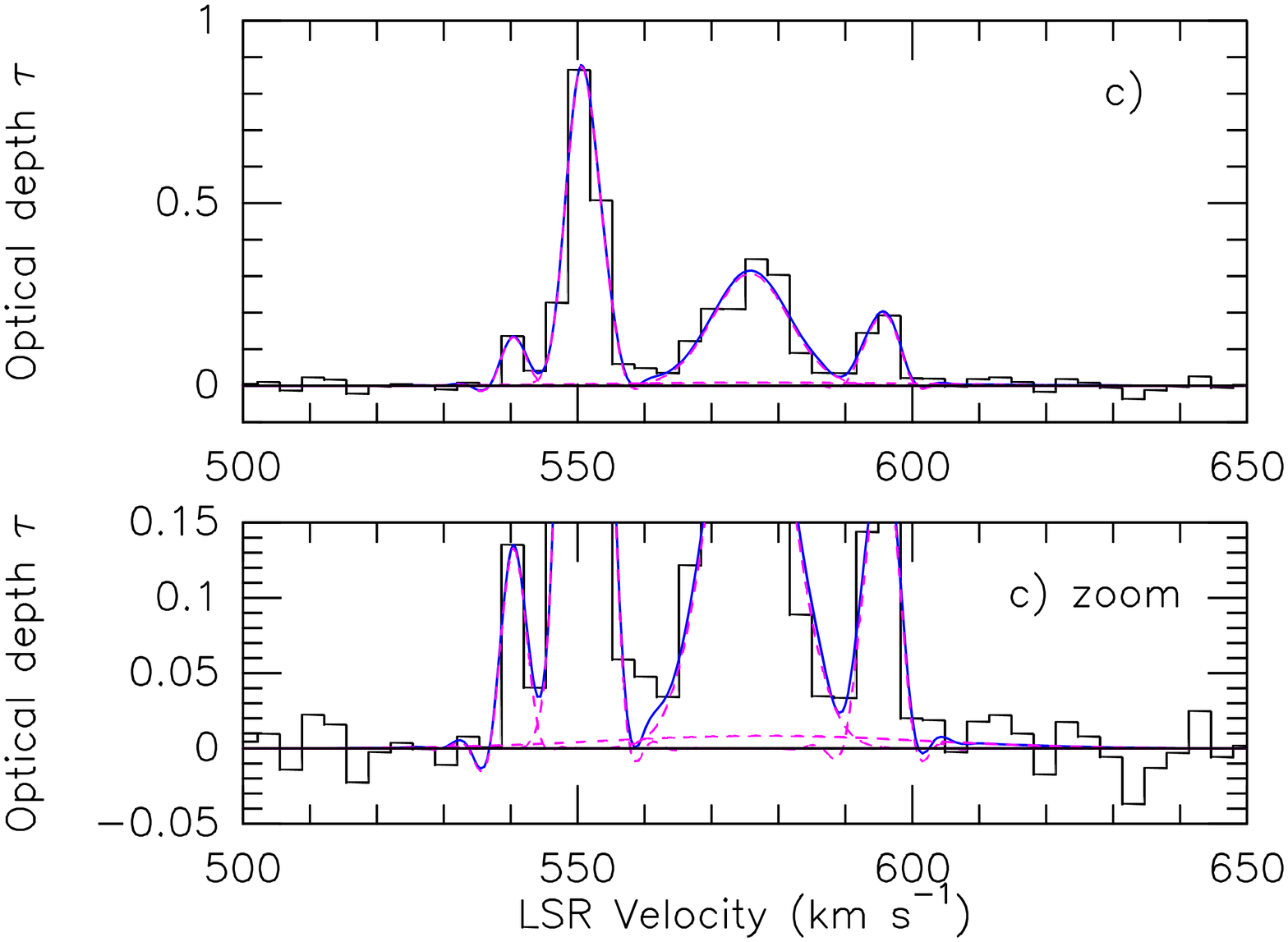}
\includegraphics[width=6.5cm,angle=-90]{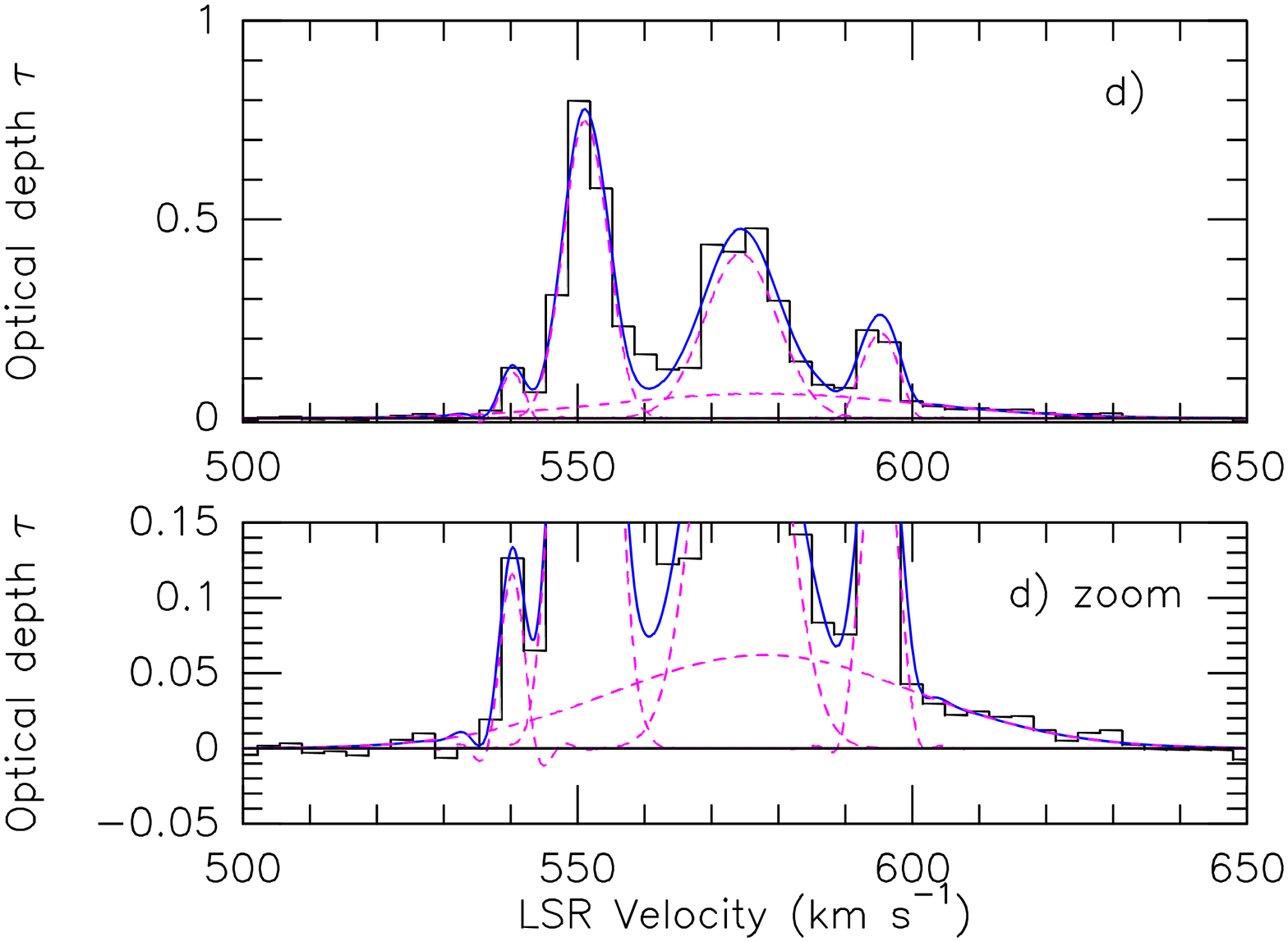}\\
\caption{ Fits to the  \ion{H}{1}\ optical depth ($\tau$) profiles at positions $c$ (left) and $d$ (right).  A zoom in optical depth from -0.05 to 0.15 is included to emphasize the weak lines. \label{fig6-5}}
\end{figure}

It is surprising that we detect the broad line component against the position $d$ but it is weaker toward the adjacent continuum region, position $c$.  
Figure~\ref{fig6-5} shows a close-up of the $\tau_{\rm HI} $ in both locations. A fit to the broad component (assuming that a Gaussian function applies) gives integrated optical depths of $\int \tau_{\rm HI}$d$V$ = 3.6 $\pm$ 1.3 \kms\ for position $d$ while it is $\int \tau_{\rm HI}$d$V$ = 0.5 $\pm$ 1.1  \kms\ in position $c$ (Table~\ref{tbl-6}). That the  broad absorption line in position $c$ is weaker is reinforced by the smaller optical depths in the spectral regions between the lines SV, I/II and III at $V$ = 560 \kms\ and 590  \kms\   with respect to position $d$ (Figure~\ref{fig6-5}).  This implies that the \ion{H}{1} opacity of the broad line component varies within a projected linear scale of less than 0.4~pc.

Finally, we calculate the \ion{H}{1} column densities (Column 7 in Tables~\ref{tbl-5} and \ref{tbl-6}), using the following relation: 
$$
N(\rm HI)/T_{\rm s}~[{\rm cm^{-2}~K^{-1}}]=  1.823 \times 10^{18} \int\tau_{\rm HI} dV ~[\rm km~s^{-1}]
$$
where $T_{\rm s}$ is the spin temperature. The \ion{H}{1} column densities in the different lines cover the range $N($\ion{H}{1}$)/T_{\rm s}$ $\sim$ (1 -- 12) $\times$ 10$^{18}$ cm$^{-2}$ K$^{-1}$. 
The broad line toward position $d$ has a corresponding $N($\ion{H}{1}$)$/$T_{\rm s}$ that is comparable to the deepest narrow lines.

\section{CO(2--1) Absorption Features}
\label{resultsCO}

 In our 1.3~mm observations, with a restoring  beam of 6\farcs0 $\times$ 2\farcs4 (or 110 $\times$ 40~pc), we 
found an unresolved continuum source at the galactic center  (Paper I).
The inner jet observed at cm wavelengths has a steep radio spectrum so the
contribution at mm wavelengths is negligible
\citep{2008A&A...483..741I}. Therefore our 1.3~mm continuum emission is dominated by the nuclear continuum emission. 
The angular size of the central continuum source is 0.5 $\pm$ 0.1~mas at 43~GHz, or 0.009 $\pm$ 0.002~pc  \citep{1997ApJ...475L..93K},  and is considerably smaller than our synthesized beam. 
 The absorption spectra of CO(2--1), $^{13}$CO(2--1) and  C$^{18}$O(2--1) (the latter two taking into account all visibilities) are shown in Figure~\ref{fig-spectrumComposition2-2}~$b)$, $c)$ and $d)$, respectively. Absorption features have been detected in CO(2--1) and $^{13}$CO(2--1), but not in C$^{18}$O(2--1)  at a 3 $\sigma$ level (= 0.3 Jy).

\subsection{CO(2--1) Line Characterization}
\label{sub:COlinechar}

The CO(2--1) absorption spectrum spans a range of 540~$<$~$V$~$<$~620~\kms\ and is composed of 6 narrow lines.  The narrow absorption features are labelled as outlined in Figure~\ref{fig-spectrumComposition2-2} $a)$: lines \emph{1},  \emph{2} and  \emph{3} correspond to the absorption lines close to the systemic velocity, at $V$ = 539, 544 and 550~\kms , and \emph{4}, \emph{5} and  \emph{6} to absorption lines at higher velocities, located at $V$ = 575, 606 and 613~\kms\ (Table~\ref{tbl-4}). Component \emph{4} is clearly detected in the profile with all visibilities, but it turns out to be weak with only the longest baselines due to the increase of noise level. 
CO(2--1) absorption line \emph{2} is located in the most likely systemic velocity of the galaxy.
Absorption lines  \emph{1},  \emph{2} and  \emph{3} are
much deeper than the higher velocity components, with  component  \emph{3} the most prominent.

Three $^{13}$CO(2--1) absorption lines are also detected, corresponding to features  \emph{1}, \emph{2}, \emph{3} in the CO(2--1) profile. An additional line corresponding to CO(2--1) line \emph{5} was tentatively detected.  Features \emph{4}  and  \emph{6} are not detected at a 3~$\sigma$ level.

The linewidths are comparable in most cases to our spectral resolution. Because of the limited velocity resolution of our SMA observations (1~\kms) and the narrowness of the absorption lines, we do not attempt to fit Gaussian functions to the several CO(2--1) line components. 
Line widths in the CO(2--1) profile of components \emph{1} and \emph{3} are about $\Delta$$V$ = 3 -- 5~\kms , while the remaining lines have even smaller line widths
$\Delta$$V$  $\simeq$ 1 -- 2~\kms.  Lines with widths less than 1 -- 2 \kms\ are not (or insufficiently)
resolved. Note that in those cases the optical depths are underestimated.

\subsection{CO(2--1) Optical Depths and Column Densities}
\label{sub:optdepthColden}

Since the radio core is small ($\sim$ 0.009~pc,  \citealt{1997ApJ...475L..93K}), the covering factor is likely close to unity.
In fact, the components close to the systemic velocity (\emph{1}, \emph{2} and \emph{3}) have depths $>$ 0.8, and assuming that $f$ $\geq$ depth of absorption for a normalized spectrum (continuum flux is unity), the covering factor is $f$ $>$ 80\%. For simplicity we adopt a filling factor $f$ of unity for all the components. 

We have calculated the CO(2--1) optical depths $\tau_{\rm CO(2-1)}$, in the same manner,
$
I _{\rm obs}[Jy] = I _{\rm cont} e^{-\tau_{\rm CO(2-1)}}
$
 (Table~\ref{tbl-4}), as those of \ion{H}{1}.
 The optical depths  for any undetected $^{13}$CO(2--1) and C$^{18}$O(2--1) lines have been calculated using a 3~$\sigma$ level upper limit.
 CO(2--1) component \emph{3} is saturated since the depth of the absorption line nearly equals the continuum level, and $\tau_{\rm CO(2-1)}$ is highly uncertain due to non-linearity of the logarithmic function.
High values of peak  $\tau_{\rm CO(2-1)} $ show that components \emph{1}, \emph{2} and \emph{3}  are optically thick. 
CO(2--1) components \emph{4}, \emph{5} and \emph{6}, as well as  all of the $^{13}$CO(2--1) lines, are characterized by peak $\tau_{\rm CO(2-1)} $  $\lesssim$ 0.8. Thus, they can be considered as optically thin if filling factor is unity.

Our CO(2--1) optical depths are reasonably well in agreement with previous single-dish measurements. The peak $\tau_{\rm CO(2-1)}$  for components \emph{1}, \emph{2} and \emph{3} are $\tau_{\rm CO(2-1)}$ = 2.1 $\pm$ 0.3, 1.6 $\pm$ 0.2 and $\geq$ 3.6 $\pm$ 1.6, respectively, which are compatible with those  obtained by \citet{1991A&A...245L..13I}, $\tau_{\rm CO(2-1)}$ = 1.8$^{+0.8}_{-0.5}$, 1.9$^{+0.8}_{-0.5}$ and $\ge$ 3.  \citet{1991A&A...245L..13I} also gives an optical depth for component \emph{4} of $ \tau_{\rm CO(2-1)}$ = 0.4 $\pm$ 0.2, while our value is 0.10 $\pm$ 0.05. The remaining lines were not detected by \citet{1991A&A...245L..13I}.
This comparison shows the need of interferometric observations in order to minimize confusion by emission and obtain higher accuracy.

As for the $^{13}$CO(2--1) lines, the peak $\tau_{\rm ^{13}CO(2-1)}$ values are:  0.11 $\pm$ 0.02, 0.19 $\pm$ 0.02, 0.58 $\pm$ 0.03 and  0.05 $\pm$ 0.01 for components \emph{1}, \emph{2}, \emph{3} and \emph{5}. \citet{1990ApJ...365..522E} gives a peak optical depth of 0.4 for the line close to $V$ = 550 \kms\  and non-detections for the rest of the lines. This value is well in agreement with our component \emph{3}.

We derive column densities of each CO(2--1),  $^{13}$CO(2--1) and C$^{18}$O(2--1) lines using the following equation:
$$
N({\rm CO}) = \frac{8 \pi \nu^3}{c^3 A_{\rm ul} g_u} \frac{ Q(k T_{\rm ex})
\exp{(E_J/k T_{\rm ex})}} {(1-\exp{(-h \nu / k T_{\rm ex})})} \int \tau_{\rm CO(2-1)} dV,
$$
\noindent where we assume that LTE conditions apply. 
$E_J$ denotes the energy of the lower level of the transition, T$_{\rm ex}$ the excitation temperature,
$Q(T_{\rm ex}$) = $\sum_{\rm J} (2J+1)\exp{(-E_{\rm J}/kT_{\rm ex})}$ the partition function, and $A_{\rm ul}$, the Einstein coefficient of each isotopologue.
We use $T_{\rm ex}$  = 10~K,
although it could vary depending on the properties and location of the molecular clouds. 
Indeed, a gradient in excitation temperature has been inferred using a nominal temperature ratio of CO(2--1) and CO(1--0) emission line ratios by \citet{1990A&A...227..342I,1991A&A...245L..13I}: $T_{\rm ex}  \simeq 10$~K at $r > 1\arcmin$ ($>1$~kpc) while it seems to be larger in the circumnuclear gas ($r < 20\arcsec$), $T_{\rm ex} \simeq 25$~K. 

The different detected lines show CO column densities $N({\rm CO})$ spanning two orders of magnitude (Table~\ref{tbl-4}), from (23 $\pm$ 3) $\times$ 10$^{14}$ cm$^{-2}$ (component \emph{4}) to $\ge$ 1530 $\times$ 10$^{14}$ cm$^{-2}$ (component \emph{3}). $N({\rm CO})$ increases approximately an order of magnitude if a $T_{\rm ex}$ = 100~K is used. 
The $^{13}$CO column density is in the range  $<$ 8 $\times$ 10$^{14}$ cm$^{-2}$ (component \emph{4}) and (145 $\pm$ 4) $\times$ 10$^{14}$ cm$^{-2}$ (component \emph{3}).

We list in Table~\ref{tbl-4} the H$_2$ column densities ($N(\rm H_2)$) assuming abundance ratios [CO]/[H$_2$] = 8.5 $\times$ 10$^{-5}$ \citep{1982ApJ...262..590F}, [$^{13}$CO]/[H$_2$] = 1 $\times$ 10$^{-6}$ \citep{1979ApJ...232L..89S} and [C$^{18}$O]/[H$_2$] = 1.7 $\times$ 10$^{-7}$ \citep{1982ApJ...262..590F}.
The H$_2$ column densities as obtained from CO(2--1) should be similar to those obtained with $^{13}$CO(2--1).  However, $N(\rm H_2)$ obtained with CO(2--1) are systematically smaller than those obtained using $^{13}$CO(2--1). The
[CO]/[$^{13}$CO] ratios  for the six lines in order of increasing velocity are: 17 $\pm$ 2, 5 $\pm$ 1, $\geq$ 10, $>$  3, 17 $\pm$ 7 and $>$  5. These abundance ratios are considerably smaller than the assumed [CO]/[$^{13}$CO] = 85, similar to the local value.
This deviation in the abundance ratios might be due to a combination of the following reasons: $i)$ different excitation temperatures for $^{12}$CO and $^{13}$CO, $ii)$ optical depth effects, $iii)$ different filling factors and $iv)$ chemistry, such as selective photodissociation (of $^{13}$CO). However, low abundance ratios of $\sim$ 25 have been observed towards the Galactic center and PKS~1830-211 \citep[see Table~7,][]{2006A&A...458..417M}

\subsection{ [$^{12}$C]/[$^{13}$C] and  [$^{16}$O]/[$^{18}$O] Isotopic Ratios}
\label{sub:isotopicratios}

We can roughly estimate the [$^{12}$C]/[$^{13}$C] and [$^{16}$O]/[$^{18}$O] isotopic ratios from the [CO]/[$^{13}$CO] and   [$^{13}$CO]/[C$^{18}$O]  ratios, assuming that the molecular abundance ratios can be decomposed into isotopic abundance ratios of the atoms. If we assume that [$^{12}$CO]/[$^{13}$CO] = [$^{12}$C]/[$^{13}$C], then we obtain that [$^{12}$C]/[$^{13}$C] $\simeq$ 5 -- 17. 
The [$^{12}$C]/[$^{13}$C] ratio is a measure of the primary to secondary processing: while $^{12}$C mostly arises from massive stars, $^{13}$C is predominantly produced from reprocessing of $^{12}$C from earlier stellar generations \citep[e.g.][]{1994ARA&A..32..191W}.

From our observations we derive lower limits from the abundance ratio [$^{13}$CO]/[C$^{18}$O]. It ranges from $>$2 to $>$12.
 This is somewhat less restrictive than the  [$^{13}$CO]/[C$^{18}$O] $>$ 30  obtained by \citet{1997A&A...322..419W} using the J=1--0 absorption line, although their spectral resolution is a factor 2 lower ($\sim$ 1.9~\kms).
 Higher spectral resolution observations, with a channel width of 0.23~\kms, had been previously carried out by \citet{1991A&A...245L..13I}. \citeauthor{1991A&A...245L..13I} report a 2~$\sigma$ detection in the C$^{18}$O line for component \emph{3}, which gives a value for the abundance ratio [$^{13}$CO]/[C$^{18}$O] $\simeq$ 13 $\pm$ 3. This would be in agreement with our lower limit, but not with that in \citet{1997A&A...322..419W}. 
 
We can then obtain  [$^{16}$O]/[$^{18}$O] from our previous estimate of [$^{12}$C]/[$^{13}$C]. 
$^{16}$O and $^{18}$O form in massive stars and are predicted to behave similarly, except at early stages of massive star formation since  $^{16}$O is a primary element and $^{18}$O is a secondary one \citep{1996A&A...309..760P}. The ratio is therefore expected to decrease with time and stellar processing \citep[e.g.][]{2006A&A...458..417M}.

By adopting [$^{12}$C]/[$^{13}$C] = 5 -- 17 and [$^{13}$CO]/[C$^{18}$O] $\simeq$ 13 $\pm$ 3 \citep{1991A&A...245L..13I} (component $3$), then  [$^{16}$O]/[$^{18}$O] $\sim$ 70 $-$ 220 . 
This ratio would be in agreement with values $\sim 150 - 200$ in nearby starburst galaxies such as NGC~253 and NGC~4945 \citep[][]{2004A&A...422..883W,1999MNRAS.303..157H,1993A&A...268L..17H}. 

Overall, the [$^{12}$C]/[$^{13}$C] and [$^{16}$O]/[$^{18}$O] ratios in Cen~A are within those found toward the Galactic center and starburst galaxies \citep{1993A&A...268L..17H}.
Thus, our calculated [$^{12}$C]/[$^{13}$C] and [$^{16}$O]/[$^{18}$O] ratios using CO lines suggest within the uncertainties that
 the molecular clouds producing the absorption lines close to the systemic velocity might have been affected by relatively long-term starburst conditions. 

On the other hand, note that  \citet{1997A&A...324...51W} derived an isotopic ratio [$^{12}$C]/[$^{13}$C] $>$ 70 for component \emph{3} from its [HCO$^+$]/[H$^{13}$CO$^+$]  ratio, which yields [$^{16}$O]/[$^{18}$O]  $>$ 200 -- 840. This seems to contradict our results and shows the need of a systematic chemical study using different isotopic molecular lines.

\section{Comparison of atomic and molecular absorption lines}
\label{subsec:comparison}

In this section we compare the variations of $\tau_{\rm HCO^+(1-0)}$/$\tau_{\rm CO(2-1)}$
and $\tau_{\rm CO(2-1)}$/$\tau_{\rm HI}$ in order to infer the variation of molecular to atomic abundances
with velocity. As a zeroth-order approximation we consider three assumptions: $i)$ absorption lines of different species (both molecular and atomic) that have similar velocities are physically related; $ii)$ temperature (and spin) differences vary similarly for different species so that the ratios are reasonably constant; and $iii)$ molecular abundance ratios are reasonably constant ([CO]/[H$_2$] and [HCO$^+$]/[H$_2$]).

\subsection{Molecular gas as traced by CO(2--1) and HCO$^+$(1--0)}

\begin{figure}
\centering
\includegraphics[width=16cm, angle=-90]{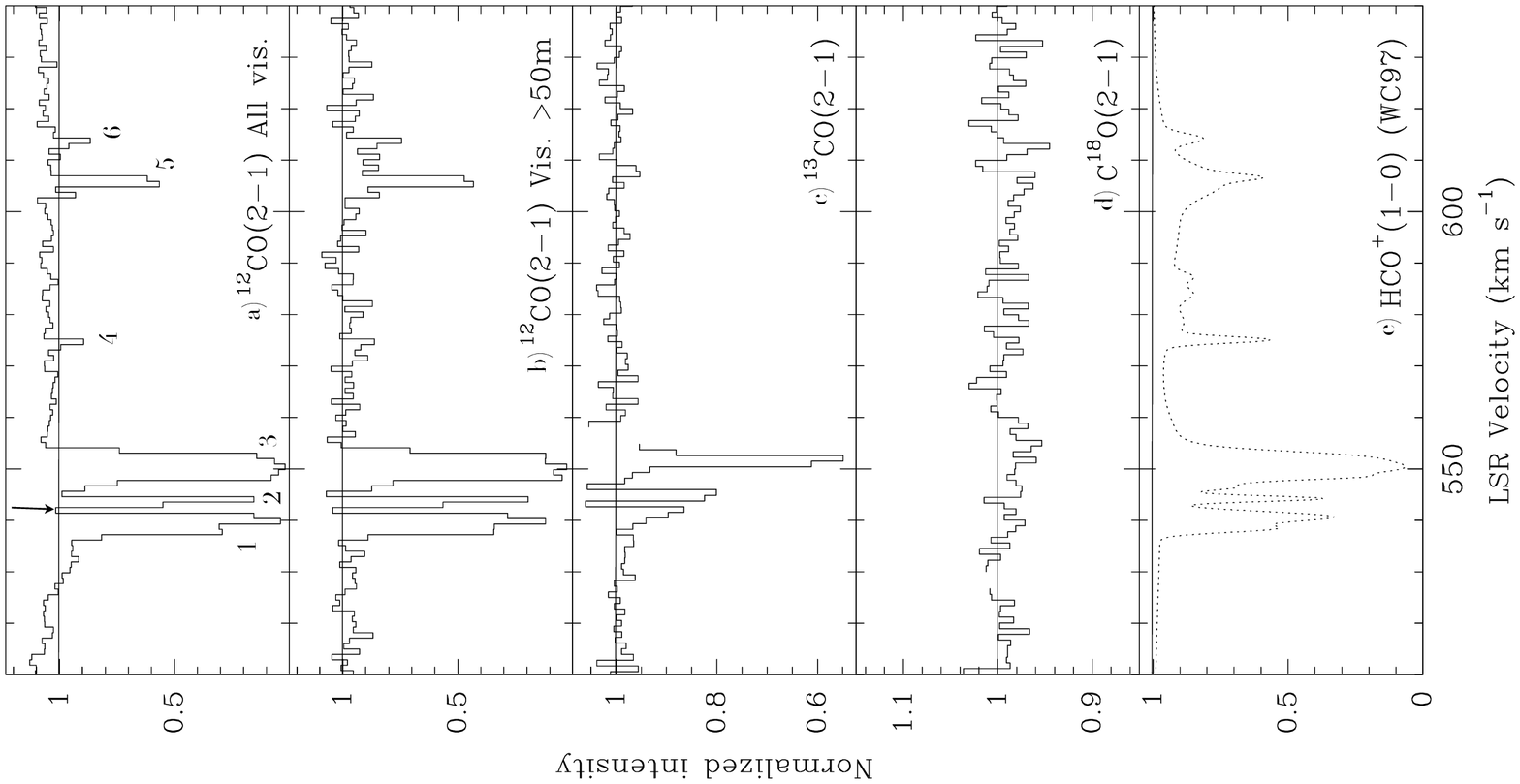}
\caption{ $a)$ CO(2--1) normalized intensity spectrum using all visibilities.  Numbered labels indicate the different molecular gas absorption components. $b)$ CO(2--1) spectrum only with projected baselines longer than 50 m. $c)$ $^{13}$CO(2--1) spectrum. The three main features close to the systemic velocity are detected.  $d)$ C$^{18}$O(2--1) spectrum. No absorption lines have been detected with our sensitivity. No emission is seen either in the $^{13}$CO(2--1) or  the C$^{18}$O(2--1) spectra, therefore we include all the visibilities.
$e)$ Gaussian fits to the single-dish HCO$^+$(1--0) spectrum \citep{1997A&A...324...51W}.
 The arrow indicates the systemic velocity ($V$ = 541.6~\kms ).
\label{fig-spectrumComposition2-2}}
\end{figure}

The different velocity components detected in the CO(2--1) absorption spectrum correspond well to other molecular lines, such as those of HCO$^+$(1--0).  We chose HCO$^+$(1--0) for comparison because it traces molecular gas with different physical conditions (its critical density is two orders of magnitude larger than that traced by CO(2--1)). It is also the best S/N and the highest spectral resolution ($\simeq$ 0.2 \kms) absorption profile that can be found in the literature \citep{1997A&A...324...51W}. The CO(2--1) absorption profile can be compared to that of HCO$^+$(1--0) (Figure~\ref{fig-spectrumComposition2-2} $e)$, which results from the sum of 17 Gaussian components used by \citet{1997A&A...324...51W} to parametrize the observed HCO$^+$(1--0) absorption spectrum. The more  prominent lines of the HCO$^+$(1--0) spectrum are centered at the same velocities and have similar widths to their CO(2--1) counterparts. 
Note that no variability was found in different epochs for the HCO$^+$(1--0) absorption spectrum \citep{1997A&A...324...51W}, so we do not expect to find variability in our CO(2--1) profiles either.

On the other hand,  the high velocity component of the HCO$^+$(1--0) absorption profile \citep{1997A&A...324...51W,1999ApJ...516..769E}  (Figure~\ref{fig-spectrumComposition2-2} $e$) is more prominent than in our  CO(2--1) spectrum (Figure~\ref{fig-spectrumComposition2-2} $b$).
The broad component of HCO$^+$(1--0) absorbs roughly 15\% of the continuum emission \citep{1997A&A...324...51W}. 
If the CO(2--1) broad absorption feature were as deep as the  HCO$^+$(1--0) broad line then we would have detected it.

We derive the abundance ratio assuming $T_{\rm ex}$ = 10~K for CO(2--1) and $T_{\rm ex}$ = 5~K   for HCO$^+$(1--0) \citep{1997A&A...324...51W}. 
For lines within 540~\kms~$V$ $<$ 570~\kms, the total abundances of HCO$^{+}$ and CO are $N({\rm HCO^+})$ $\simeq$ 4.8 $\times$ 10$^{13}$ cm$^{-2}$ \citep[Table~8,][]{1997A&A...324...51W} and $N({\rm CO})$ $\simeq$ 2.3 $\times$ 10$^{17}$  cm$^{-2}$, which provides the ratio  [HCO$^{+}$]/[CO] $\simeq$ 2 $\times$ 10$^{-4}$. In the high velocity component, $V$ $>$ 570~\kms, the ratio is five times larger than in the low velocity components, [HCO$^{+}$]/[CO] $\simeq$ 1 $\times$ 10$^{-3}$.

The [HCO$^{+}$]/[CO]  ratio is even larger within 570~\kms\ $<$ $V$ $<$ 600~\kms . 
While [HCO$^{+}$]/[CO] $\simeq$ 5 $\times$ 10$^{-4}$ for components at $V$ $>$ 600~\kms, in the velocity range between 570~\kms\ $<$ $V$ $<$ 600~\kms\ it is a factor 10 larger, [HCO$^{+}$]/[CO] $\simeq$ 5 $\times$ 10$^{-3}$.
This difference arises in the velocity range corresponding to the broad absorption component.

\subsection{Atomic versus molecular gas content}
\label{subsec:comparaHICO}

\begin{figure}
\centering
\includegraphics[width=7cm,angle=-90]{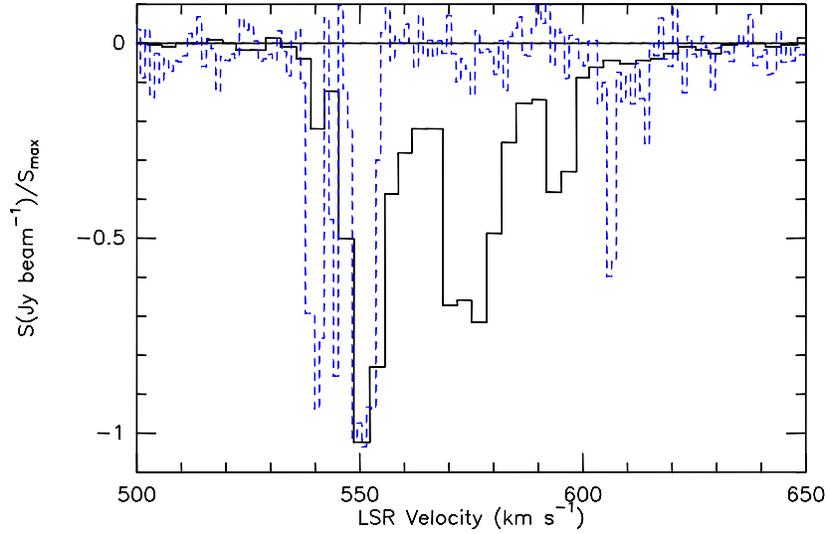}
\caption{Comparison of the $^{12}$CO(2--1) (dashed line) and the \ion{H}{1}\ absorption profiles (solid line). The velocity resolution is 1 \kms\ and 3.3 \kms\ for CO(2--1) and \ion{H}{1} , respectively. \label{fig6-3a}}
\end{figure}

In Figure~\ref{fig6-3a}  we present a comparison between the \ion{H}{1} and   CO(2--1) absorption features. 
It is important to note that we are tracing  both the \ion{H}{1} component $d$ and molecular gas in the line of sight within a  region located closeby ($\lesssim$ 0.3~pc).

A wide range of atomic to molecular abundance ratios are seen.  
While the main \ion{H}{1}  absorption features are located between $V$ = 544 and 600 \kms, in CO and other molecules the main absorption features occur close to the systemic velocity, and then at larger velocities, $V$ $>$ 605 \kms.
The \ion{H}{1}\ absorption features are remarkably more prominent at 560~$<$~$V$~$<$~610 \kms\ than their molecular counterparts. The two components \emph{5} and \emph{6} of the CO(2--1) and HCO$^+$(1--0) spectra at $V$ $>$ 605 \kms\ are located in the same velocity range as the \ion{H}{1} broad absorption feature which is seen up to $V$ = 640 \kms . 

Next we compare the molecular to atomic abundance ratio assuming $T_{\rm ex}$ = 10~K for CO(2--1) and $T_{\rm s}$= 100~K for \ion{H}{1}. 
First, in the low velocity component the gas is mostly in its molecular phase.
The CO(2--1) line \emph{1} seems to be related to the \ion{H}{1} component IV, and it yields a ratio $N({\rm H_2})$/$N($\ion{H}{1}$)$ = (7.25 $\times$ 10$^{20}$ cm$^{-2}$) / (0.9 $\times$ 10$^{20}$ cm$^{-2}$)  $\simeq$ 8.  
 The $N({\rm H_2})$/$N($\ion{H}{1}$)$ ratio corresponding to CO(2--1) line \emph{2} is similar to that of the CO(2--1) line \emph{1}.
The \ion{H}{1} absorption line SV and CO(2--1) line \emph{3}, the ratio of H$_2$ to \ion{H}{1}\ column densities is $N({\rm H_2})$/$N($\ion{H}{1}$)$ $\geq$ ($1800 \times 10^{18}$~cm$^{-2}$) / ($12 \times 10^{20} $ cm$^{-2}$) $=$ 1.5.

In the high velocity component located between 570~\kms $<$ $V$ $<$ 600~\kms, the gas is predominantly either atomic or molecular gas traced by HCO$^+$(1--0), but not by CO(2--1). \ion{H}{1} lines I/II and III contain a total of $N($\ion{H}{1}$)$ $\simeq (13.4 \pm 2) \times 10^{20}$~cm$^{-2}$. If spatially linked to molecular clouds with the same kinematics, they would only correspond to CO(2--1) line \emph{4}, which contains a smaller amount of molecular gas, $N({\rm H_2})$ $\simeq (26 \pm 4) \times 10^{18}$~cm$^{-2}$.  The ratio is $N({\rm H_2})$/$N($\ion{H}{1}$)$ $\simeq$ 0.02, 2 -- 3 orders of magnitude smaller than in the low velocity components. However, we have to add the contribution from the molecular gas column density traced by other coexisting species such as HCO$^+$(1--0) in this velocity range. \citet{1997A&A...324...51W} estimated it to be $N({\rm HCO^+})$/$N($\ion{H}{1}$)$ $\simeq$ 1.5 $\times$ 10$^{-8}$, or $N({\rm H_2})$/$N($\ion{H}{1}$)$ $\simeq$ 0.2 -- 2 assuming a typical  $\rm [HCO^+]$/$[\rm H_2]$ = 10$^{-7}$ -- 10$^{-8}$. Even with the contribution of HCO$^{+}$ in the velocity range between 570~\kms $<$ $V$ $<$ 600~\kms , we obtain a slightly smaller $[\rm H_2]$/[\ion{H}{1}] ratio than in the low velocity component. 

Finally, from $V$ $>$ 600~\kms ,  CO(2--1) components \emph{5} and \emph{6} are within the \ion{H}{1} broad line. There $N($\ion{H}{1}$)$ = $ 7 \times 10^{20} $ cm$^{-2}$ (broad line), to be compared to the $N({\rm H_2})$ $\simeq 2.6 \times 10^{20}$~cm$^{-2}$, which yields a ratio $N({\rm H_2})$/$N($\ion{H}{1}$)$ $\simeq$ 0.4.

\section{Discussion }
\label{discussion}

\subsection{Warped Disk or Non-Circular / Infalling Motions?}
\label{models}

Although there is a consensus that narrow absorption lines with velocities close to the systemic velocity (Low Velocity Component -LVC-, in the velocity range 540 \kms  $<$ $V$ $<$ 560 \kms , following the nomenclature by \citealt{1997A&A...324...51W}) represent gas far from the nucleus ($\sim$ 1~kpc), still there is no agreement regarding the more redshifted components (High Velocity Component -HVC-, 560 \kms  $<$ $V$ $<$  640 \kms ).

\begin{figure}
\centering
\includegraphics[width=12cm,angle=90]{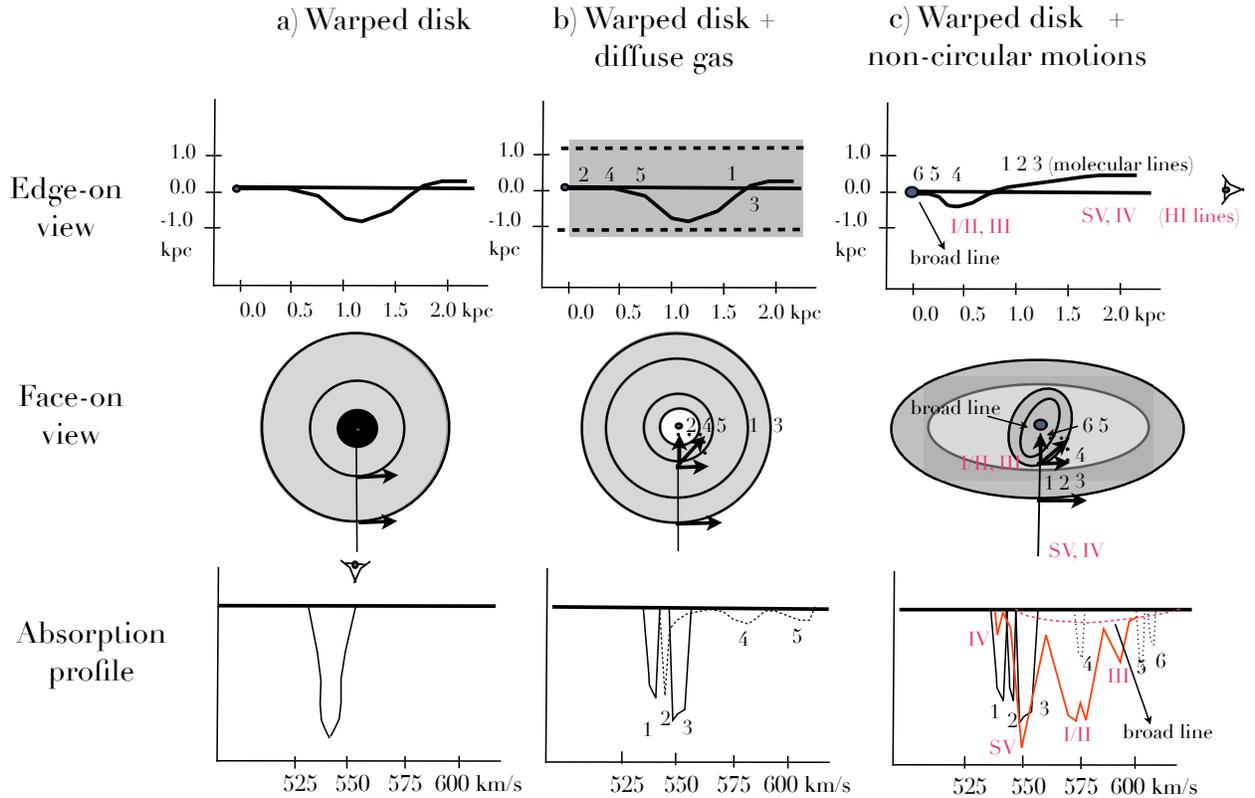}
\caption{Scheme showing the edge-on and face-on views of the proposed models within the inner 2~kpc, as well as the corresponding absorption lines: $a)$ warped thin disk (as in Quillen et al. 2005), $b)$ warped disk with diffuse gas \citep{1999ApJ...516..769E}, and $c)$ warped disk with a contribution of non-circular motions (weak bar) model \citep{2009ApJ...695..116E}.} \label{figscheme}
\end{figure}

A schematic view of proposed models to describe the almost edge-on disk of CenA is presented in Fig.~\ref{figscheme}. This figure includes edge-on and face-on views of the proposed models within the inner 2~kpc, as well as a qualitative interpretation of the kinematic location of the absorption lines.
A warped and thin disk model \citep[e.g.][]{2006ApJ...645.1092Q,2007ApJ...671.1329N,2009arXiv0912.0632Q} 
reproduces well the observed emission lines in terms of distribution and kinematics. However, this model is not able to reproduce  by itself  the existence of the redshifted components (HVC) because it assumes pure circular orbits \citep{1999ApJ...516..769E} (see Figure~\ref{figscheme} $a$). 
It is necessary to augment this model with either diffuse gas \citep{1999ApJ...516..769E} (see Figure~\ref{figscheme} $b$) or non-circular/infalling motions (e.g., as suggested in \citealt{1983ApJ...264L..37V} and Paper I;  see Figure~\ref{figscheme} $c$).
Since the implications of the latter two models diverge, we discuss next what is the most plausible one.

It is remarkable that we detect the broad line of the HVC against the brightest continuum source (position $d$) but it is found to be much weaker toward position $c$ just $\sim$ 0.4~pc next to it and further from the nucleus. This considerably restricts the upper limit of a few hundred parsecs set by VLA data \citep{2002ApJ...564..696S}.  The non-circular/infalling motions within a warped disk model (Figure~\ref{figscheme} $c$) can provide the observed differences between positions $c$ and $d$, since one would expect higher velocities closer to the nucleus. On the other hand, this is not expected in the  warped disk with diffuse gas model (Figure~\ref{figscheme} $b$) since this gas should be distributed at high latitudes up to about a few hundreds of parsecs.

In addition, the physical locations of the different absorbing regions in the line of sight derived from  the warped disk with diffuse gas model (Figure~\ref{figscheme} $b$) are complex. The model predicts that LVC lines \emph{1} and \emph{3} (Figure~\ref{fig-spectrumComposition2-2}) arise at $r$ = 1.7 -- 1.9~kpc and are separated by up to $\Delta z$ $\simeq$ 160~pc from the disk, line \emph{2}  should be located $r$ = 200~pc and $\Delta z$ $\simeq$ 0,  and the HVC lines \emph{4}, \emph{5} and \emph{6} at $r$ = 200 -- 600~pc  and  $\Delta z$ $\simeq$ 300~pc  (see Figure~\ref{figscheme} $b$).
In this model, within $r$ $<$ 600~pc, the nearer the corresponding absorbing region is to the nucleus the closer the corresponding line velocity is to the systemic velocity.

Finally, no component within the inner $r$ = 200~pc is taken into account in Eckart et al.'s model, although a large concentration of molecular gas is found there (Paper I).  A contribution to the molecular gas profile of this component should be seen as an extension of the narrow line \emph{2} (arising from $r$ = 200~pc at $\Delta z$ = 0~pc in \citeauthor{1999ApJ...516..769E}'s model). 
The limit imposed by \citet{1999ApJ...516..769E} was set at $r$ = 200~pc so that the line \emph{2} reproduces the observed one. However, without that constraint, line \emph{2} is difficult to be reproduced in that model.

These arguments suggest that the nature of the HVCs is not just diffuse gas within a warped disk far from the nucleus, and that non-circular motions and/or infalling gas must be important mechanisms to explain the nature of this component. Within this scenario, molecular lines $4$, $5$ and $6$ and atomic lines I/II and III would arise at radii $r$ $\lesssim$ 1~kpc, while lines $1$, $2$ and $3$ and SV and IV would be at  $r$ $\gtrsim$ 1~kpc (Figure~\ref{figscheme} $c$).

\subsection{An X-ray Dominated Region (XDR)?}

Since a powerful AGN resides in this galaxy, and a significant amount of gas is located within the inner $r$ = 220~pc (Paper I), it is reasonable to think that X-ray radiation plays a major role modifying the chemical complexity (i.e., dissociating molecular hydrogen into atomic gas, ionizing and heating, etc.).

In fact, the AGN of Centaurus A is characterized by a  X-ray luminosity of $L_{\rm X}$ $\sim$ 4.8 $\times$ 10$^{41}$~erg s$^{-1}$. The incident flux at a distance of 1~pc is thus quite large, $F_{\rm X}$ $\sim$ 300~erg s$^{-1}$ cm$^{-2}$. On the other hand, it is not likely that it has such a strong effect at distances $\sim$100~pc, with fluxes of the order of $F_{\rm X}$ $\sim$ 0.03~erg s$^{-1}$ cm$^{-2}$. 
The gas column density attenuating these X-rays is estimated to be  $\sim$ 10$^{23}$ cm$^{-2}$,  probably by a cloud that entirely surrounds the nucleus or by a torus-like structure \citep[][ and references therein]{2004ApJ...612..786E}. 

The LVC is likely arising from cold molecular gas, with kinetic temperatures $T$ $\sim$ 10~K and average hydrogen densities of the order of $n$ $\sim$ 10$^{4}$ cm$^{-3}$,  which seems to be appropriate for the disk as seen in emission \citep[e.g.][]{1990ApJ...363..451E} as well as in absorption  \citep[e.g.][]{1997A&A...324...51W,2009ApJ...696..176M}.
Less information is present in the literature for the physical conditions of the HVC, since this weak component is difficult to detect. 

Observationally, \citet{1997A&A...324...51W} show that the average of [HCO$^+$]/[HCN] $\sim$ 1.3 and 1.7, while the average for [HNC]/[HCN] is  0.25  and  0.5, respective to the LVC and the HVC. 
The numerical ratios of \citet{2005A&A...436..397M} for a XDR are  [HCO$^+$]/[HCN] $\sim$ 1 -- 10 and [HNC]/[HCN] $\sim$ 0.01 -- 1, with  $F_X$=160~erg s$^{-1}$ cm$^{-2}$ and $n_H$ = 10$^{5.5}$ cm$^{-3}$ (or both $F_X$ and $n_H$ two orders of magnitude lower). This numerical ratio agrees well with the observations.
However, under certain circumstances, a PDR could also reproduce the given molecular gas ratios. 

X-rays do not lead to strong dissociation of CO and it can be present in an XDR at elevated temperatures \citep{2008EAS....31...47S}.
Warm CO gas produces emission originating from high rotational transitions. Bright CO emission lines have been seen for the first three J levels, and larger $T_{ex}$ (20 -- 30~K) have been inferred for molecular clouds close to the nucleus \citep[e.g.][]{1991A&A...245L..13I}. 
On the other hand, absorption lines  of the lower J transitions will preferentially sample the (excitationally) coldest gas.
In the nuclear regions, where higher densities and kinetic temperatures are expected, only high-J transitions of CO would have such broad lines as seen in HCO$^{+}$(1--0), as it has been suggested with previous CO(3--2)  single-dish data \citep{1997A&A...324...51W}. This is likely the reason why we did not detect the broad line in CO(2--1), but may well be detected in higher transitions of CO. This could be a good test to distinguish the physical condition of the molecular gas closer to the nucleus.

\label{physicalHI}

The molecular lines  at $V$~$>$~605 \kms , including absorption features \emph{5} and \emph{6}, are probably molecular clumps close to or within the \ion{H}{1} region from which the \ion{H}{1} broad absorption component arises.
We consider the possibility that these features are located close to the nucleus (as suggested in Figure~\ref{figscheme} $c$).
For an X-ray illuminated gas like these clouds, the physical conditions of the gas that surrounds the AGN can be studied using the effective ionization parameter, $\xi_{\rm eff}$, which is proportional  to the ratio of incident (and attenuated) X-ray photon flux to gas density of the cloud  \citep{1996ApJ...466..561M}. The effective ionization parameter of a molecular cloud can be calculated as:

$$
\xi_{\rm eff} \sim 1.1 \times 10^{-2} L_{44}/ (N_{22}^{0.9} n_9 r^{2})
$$

\noindent where the $L_{44}$ = $L_{\rm X}$/10$^{44}$ is the X-ray luminosity of hard photons (energies $>$ 2 keV) in ergs s$^{-1}$, $n_9$  the gas density of the cloud in 10$^9$ cm$^{-3}$, $r$ the distance to the X-ray emitting source in pc, and $N_{22}$ the attenuating gas column density in 10$^{22}$ cm$^{-2}$ \citep{1994ApJ...432..606M,1996ApJ...466..561M}.
The molecular cloud regions are exposed to an X-ray luminosity from Cen~A's AGN of 
$L_{\rm X}$ = 4.8 $\times$ 10$^{41}$ ergs s$^{-1}$ in the hard energy range 2 --10 keV \citep{2004ApJ...612..786E}. 

If we assume for the molecular gas arising in the two high velocity absorption features \emph{5} and \emph{6} a T$_{\rm ex}$ = 10~K, a size of the 1.3~mm continuum of 0.01~pc \citep{1997ApJ...475L..93K}, and a size of the cloud of R $\simeq$ 10~pc then the molecular clouds would be characterized by densities of about $n$ $\sim$ 10$^{5}$ cm$^{-3}$.

The gas column densities that are attenuating the X-rays are estimated  to be  $\sim$ 10$^{23}$ cm$^{-2}$ \citep{2004ApJ...612..786E}. 
By assuming a distance to the nucleus of $r$ = 10~pc, then it yields a parameter $\xi_{\rm eff}$ = 7 $\times $10$^{-4}$. For 
$r$ = 1~pc then $\xi_{\rm eff}$ = 0.1. A $\xi_{\rm eff}$ $>$ 10$^{-3}$ means that the gas will be predominantly in its atomic phase and it indicates that the molecular clouds must be at $r$ $>$ 10 pc in order to survive under the above assumptions. This is in agreement with the smaller [H$_2$]/[\ion{H}{1}] ratio found in the HVC with respect to the LVC (\S~\ref{subsec:comparaHICO}).

\subsection{Physical properties of the circumnuclear \ion{H}{1}}

As mentioned in \S~\ref{sub:HIabsorptionfeatures} the velocity widths of our three main prominent \ion{H}{1} features (i.e. I, I/II and III) are between 5 -- 10~\kms .  If one neglects line broadening due to rotation and turbulence, and attributes the observed FWHM only to thermal motions in the gas, the line widths shown in  Table~\ref{tbl-5} would indicate temperatures between 800 and 5,700~K.
Turbulence can account for a linewidth of about $\sim$ 5 \kms , at least within our own Galaxy  \citep{1988gera.book..295B}. Taking this into account, the actual gas temperature is probably between 20 and 2500~K.

The nature of the gas from which the broad absorption lines arise are much less understood.
The underlying broader \ion{H}{1} absorption line, with a width of $\sim$ 55~\kms , cannot be attributed solely to thermal motions since it would give too high temperatures ($\sim$ 60,000~K). As indicated by  \citet{2002ApJ...564..696S} this redshifted  broad \ion{H}{1} component is probably a blend arising from several complexes rather than a single one. Since it only shows redshifted velocities, it is likely that this component is largely affected by systematic kinematic contributions, such as non-circular and/or infalling motions, and close to the nucleus, mostly within a distance $\lesssim$ 0.7~pc (0.4~pc from position $c$ to $d$, and 0.3~pc from position $d$ to the actual nucleus; as mentioned in \S~\ref{21cmcontinuum}).
 If the geometry of the absorbing atomic gas is similar to the molecular gas disk/torus in Paper I (inclination  $i$ = 70$\arcdeg$), then the disk-like feature associated with the broad \ion{H}{1} line with column densities of about $N($\ion{H}{1}$)$ $\sim$ 7 $\times$ 10$^{20}$~cm$^{-2}$ would have a (non-projected) radius $r$ $\lesssim$ 2~pc. However, we cannot rule out other geometries.  
Also, note that the kinematic coincidence between the broad absorption lines of \ion{H}{1} and other molecular lines such as HCO$^+$(1--0) may suggest that the corresponding absorbing molecular regions are physically connected.

The atomic material arising from the broad component is likely contributing to fuel the powerful AGN of Centaurus A. Assuming that this \ion{H}{1} region is infalling with a velocity of $\sim$ 37~\kms (mean velocity minus the systemic velocity), a column density of $N$(\ion{H}{1}) $\sim$ 7 $\times$ 10$^{20}$~cm$^{-2}$, and a size of 2~pc, we estimate an accretion rate of 2 $\times$ 10$^{-3}$ M$\odot$~yr$^{-1}$, enough to power the required accretion rate of $\sim$ 1.5 $\times$ 10$^{-5}$ M$\odot$~yr$^{-1}$ obtained from its isotropic radio luminosity \citep[e.g.][]{1989AJ.....97..708V}.

\section{Summary and conclusions} \label{conclusion}

We present   \ion{H}{1} and CO(2--1) absorption features toward the nuclear regions of Centaurus A (NGC 5128), using the VLBA and the SMA respectively. 

Our \ion{H}{1} observations allow us to discern with sub-pc resolution the absorption profiles toward the nuclear jet in the inner 5.4~pc. 
The  \ion{H}{1} absorption lines are composed of a system of 4 main narrow lines as well as an underlying broad line (FWHM $\sim$ 55~\kms), in agreement with previous works using lower spatial resolution observations.
Interestingly, the underlying broad absorption line in our data is seen to be more prominent toward the central 21~cm continuum component (base of the nuclear jet, although not the nucleus itself) than toward the continuum emission at $\geq$ 0.4~pc from it and further from the nucleus.

We interpret this as a broad velocity component that arises from an absorbing region that is physically close to the AGN itself, and not just as diffuse gas seen in projection  as suggested by \citet{1999ApJ...516..769E} since the height of such diffuse component is expected to be of the order of a few hundreds of parsec, and thus we would not expect any substantial difference between both positions at a 0.4~pc scale.
In addition, the large width of the broad absorption line suggests that the absorbing region is composed of several complexes with systematic non-circular or infalling motions.

As for the molecular gas component, our CO(2--1) interferometric data provides for the first time CO absorption features against the unresolved 1.3~mm continuum emission with minimal contamination by line emission. 
The narrow lines (likely arising from clouds at large radii from the nucleus) observed in the CO(2--1) and $^{13}$CO(2--1) profile were identified with previously observed molecular lines such as HCO$^+$(1--0) \citep[e.g.][]{1997A&A...324...51W,1995ApJ...448L.123V}. Overall, our calculated isotopic ratios [$^{12}$C]/[$^{13}$C] $\sim$ 5 -- 17 and the [$^{16}$O]/[$^{18}$O]  $\sim$ 70 -- 220  seem to indicate that the molecular clouds producing the observed narrow absorption lines close to the systemic velocity have been in a relatively long-term starburst environment as in nearby starburst galaxies. We did not detect any counterpart of the broad absorption line in the CO(2--1) spectrum as that found in \ion{H}{1} or other molecular lines, although the noise level in our CO(2--1) profile would have been sufficient to detect this feature if it were as strong as that of HCO$^+$(1--0).

Since both our \ion{H}{1} (position $d$) and molecular absorption profiles arise from molecular clouds in front of a similar continuum emission region within $\sim$ 0.3~pc, we can compare the atomic and molecular gas properties along the line of sight.  
The abundances of \ion{H}{1} to H$_2$ are noticeably different at different velocities. 
While the low velocity components, $V$ $<$ 560~\kms\, are dominated by molecular gas, the atomic phase gas seems to dominate at 560 $<$ $V$ $<$ 600~\kms\  when the \ion{H}{1} line is compared with the molecular gas content as traced by the CO(2--1) line. 
Although the molecular to atomic gas ratio is not drastically different when we consider the molecular gas traced by HCO$^+$(1--0), still there is a trend for a larger atomic content for components at higher velocities.  
These distinct signatures suggest that the physical properties, location and chemistry of the gas producing the broad line is different to those of the narrow lines. 

Since the gas corresponding to the broad line is likely not far from the AGN, we argue that its properties might be modified by X-rays, which penetrate much farther than UV. The observed abundances of species such as HCO$^+$ and HCN were estimated to be $\sim$ 10$^{-8}$ , and abundance ratios [HCO$^+$]/[HCN] $\geq$ 1 and [HNC]/[HCN] $\sim$  0.5, which seems to agree with the chemistry of a X-ray dominated region (XDR). 
In the nuclear regions, higher densities and kinetic temperatures are expected. Only high-J transitions would have such broad lines as seen in HCO$^{+}$(1--0). Since X-rays do not strongly dissociate CO molecules and they can coexist at high temperatures, higher transitions of CO other than CO(2--1) could trace the broad line component, as has been suggested with previous CO(3--2)  single-dish data.

Because of its kinematics, the two narrow red-shifted lines at $V$ $>$ 600~\kms\ found in our CO(2--1) spectrum, as well as in previous HCO$^+$(1--0) spectrum might be close to the AGN and falling toward it. These molecular clouds are likely at a distance $\gtrsim$ 10~pc from the AGN so that they can survive the X-ray emission. On the other hand, at distances $\lesssim$ 10~pc most of the gas will be in its atomic phase, material that is likely contributing to power the AGN of Centaurus A.

\acknowledgments{
We thank the SMA and NRAO staff members who made the observations reported here possible. We also thank A. Sarma for providing the 21~cm continuum VLA data, and S. Martin and I. Jimenez-Serra for interesting discussions.  This research has made use of NASA's Astrophysics Data System Bibliographic Services, and has also made use of the NASA/IPAC Extragalactic Database (NED) which is operated by the Jet Propulsion Laboratory, California Institute of Technology, under contract with the National Aeronautics and Space Administration. DE was supported by a Marie Curie International Fellowship within the 6$\rm ^{th}$ European Community Framework Programme (MOIF-CT-2006-40298).}

{\it Facilities:} \facility{VLBA,SMA}

\bibliography{cena-absorption}

\begin{table}
\begin{center}
\caption{Main parameters of the VLBA   \ion{H}{1} and SMA CO(2--1) observations \label{tbl-2}.}
\begin{tabular}{ll}

\tableline
{\bf   \ion{H}{1}\ VLBA observations}&	                     \\
\tableline
& \\
Date  & 1994 November 26 \\
R.A. of phase center (J2000)   & 13${\rm ^h}$25${\rm ^m}$27.${\rm^s}$609  	\\
Decl. of phase center (J2000)   &  --43\arcdeg01$\arcmin$07$\farcs$97       \\
Time on source (hr)  &  2 \\
FWHM of synthesized beam       & 36.3 $\times$ 7.2~mas (0.65 $\times$ 0.13~pc), PA = 14\fdg8 \\
LSR velocity at band center (\kms)&	541.6 \\
Total bandwidth & 2 MHz  ($\sim$ 430 \kms )\\
Spectral resolution (\kms)& 3.3\\
rms noise (3.3 \kms) & 3 mJy beam$^{-1}$ \\
rms (continuum map) & 2 mJy beam$^{-1}$ \\
\\\hline

{\bf CO(2--1) SMA observations}   &	                     \\\tableline
& \\
Date  &  2006 April 5 \\
R.A. of phase center (J2000)   & 13${\rm ^h}$25${\rm ^m}$27.${\rm^s}$6 	\\
Decl. of phase center (J2000)   &   --43\arcdeg01$\arcmin$09$\arcsec$     \\
On source time (hr)  &  3  \\
FWHM of synthesized beam      & 2\farcs4 $\times$ 6\farcs0 (43  $\times$ 108~pc), PA = 0\fdg82 \\
LSR velocity at band center (\kms)&	550 \\
Total bandwidth &	2~GHz ($\sim$ 2870 \kms ) in each sideband\\
               & (separated by 10~GHz)\\
Spectral resolution (\kms)& 1\\
rms noise (1 \kms) & 0.1~Jy~beam$^{-1}$ (0.3 Jy beam$^{-1}$ for baselines $>$ 50~m)\\
\tableline
\end{tabular}
\end{center}
\end{table}

\renewcommand{\baselinestretch}{1.2}
\scriptsize

\begin{table}
\begin{center}
\caption[Gaussian Functions fitted to Absorption Profile]{Parameters of \ion{H}{1}\ absorption lines integrated over the entire 21~cm continuum \label{tbl-5}}
~~~\\
\begin{tabular}{crccccc}
\tableline\tableline
Absorption & $V$   & Peak $S$ \tablenotemark{(a)} & FWHM\tablenotemark{(a)} & Peak $\tau$ & $\int \tau$d$V$\tablenotemark &$N(\rm   HI)$/$T_{\rm s}$ \\ 
feature & (\kms)& (mJy beam$^{-1}$)  & (\kms) & & (\kms) & (cm$^{-2}$ K$^{-1}$) \\
\tableline
IV       & 541 $\pm$ 1 &  $  -57 \pm 19$ &  5   $\pm$   2    & 0.13 $\pm$ 0.03 & 0.7 $\pm$ 0.4 &  $(1.2 \pm 0.7) \times 10^{18}$ \\ 
SV     & 552 $\pm$ 1 & $-323 \pm 20$ &  8.5 $\pm$  1.2 & 0.74 $\pm$ 0.03 & 7.7 $\pm$ 1.0 & $(14   \pm 2)   \times 10^{18}$ \\
I, II      & 575 $\pm$ 1 & $-203 \pm 17$ & 12.6 $\pm$ 1.6 & 0.48 $\pm$ 0.02 & 7.6 $\pm$ 0.9 & $(14   \pm 2)   \times 10^{18}$ \\
III       & 595 $\pm$ 1 & $-122 \pm 20$ &  5.2 $\pm$   1.4 & 0.25 $\pm$ 0.03 & 1.8 $\pm$ 0.4 &  $(3.2 \pm 0.7) \times 10^{18}$ \\
Broad& 578 $\pm$ 4 &  $ -34 \pm 21$ &  53   $\pm$  10   & 0.07 $\pm$ 0.03 & 4.0 $\pm$ 1.9 &  $(7 \pm 3)   \times 10^{18}$ \\
\tableline
\end{tabular}
\end{center}
\tablenotetext{a}{Flux density peak and FWHM are calculated from a Gaussian fit.
}

\end{table}

\renewcommand{\baselinestretch}{1.2}
\scriptsize
\begin{table}
\begin{center}
\caption[ Gaussian Functions fitted to Absorption Profile]{ Parameters of \ion{H}{1}\ absorption lines toward individual positions \tablenotemark{(a)}   \label{tbl-6}}
~~~\\

\begin{tabular}{crclllr}
\tableline\tableline

Absorption & $V$ &  Peak $S$ & FWHM & Peak $\tau$  & $\int \tau$d$V$&$N$(\ion{H}{1})/$T_{\rm s}$ \\ 
feature & (\kms)& (mJy beam$^{-1}$)  & (\kms) & & (\kms)& (cm$^{-2}$ K$^{-1}$) \\
\tableline
a)  \\
IV     & (541)        & $<$ 2       &  (5)        & $<$ 0.6       & $<$  2    & $<$ 4 $\times$ 10$^{18}$ \\
SV$^{c}$    &  552 $\pm$ 1 & $-2 \pm 3$ & 4 &   $<$ 0.6  & $<$  2  & $<$ 4 $\times$ 10$^{18}$ \\ 
I/II   & (575)        & $<$ 2       & (14)        & $<$ 0.6       & $<$  3    & $<$ 11 $\times$ 10$^{18}$ \\
III    & (595)        & $<$ 2       &  (6)        & $<$ 0.6       & $<$  2    & $<$ 5 $\times$ 10$^{18}$  \\
Broad & (578)        & $<$ 2       & (50)        & $<$ 0.6       & $<$ 6    & $<$ 43 $\times$ 10$^{18}$ \\
\tableline
b) \\
IV     & (541)        & $<$ 17       &  (5)        & $<$ 0.3         & $<$  1    & $<$ 2 $\times$ 10$^{18}$ \\
SV    &  552 $\pm$ 1 & $9 \pm 2$ &   5 $\pm$ 3 & 0.81 $\pm$ 0.07 & 4 $\pm$ 2 & (7 $\pm$ 3) $\times$ 10$^{18}$ \\ 
I/II   & (575)        & $<$ 17       & (14)        & $<$ 0.3         & $<$  2    & $<$ 7 $\times$ 10$^{18}$ \\
III    & (595)        & $<$ 17       &  (6)        & $<$ 0.3         & $<$  1    & $<$ 4 $\times$ 10$^{18}$ \\
Broad & (578)        & $<$ 17       & (50)        & $<$ 0.3         & $<$ 4    & $<$ 30 $\times$ 10$^{18}$ \\

\tableline
c) \\
IV     & 541  $\pm$ 1 & $129 \pm 5$ &  5   $\pm$  2   & 0.12 $\pm$ 0.02 & 0.7 $\pm$ 0.3 &  (1.3 $\pm$ 0.5) $\times$ 10$^{18}$ \\ 
SV    & 551        $\pm$ 1 & $59 \pm 5$ &  6.3 $\pm$  1.0 & 0.90 $\pm$ 0.02 & 5.5 $\pm$ 1.0 & (10   $\pm$ 2) $\times$ 10$^{18}$ \\
I/II   & 576   $\pm$ 1 & $108 \pm 5$ & 13   $\pm$  2   & 0.29 $\pm$ 0.02 & 4.4 $\pm$ 0.7 &  (8.0 $\pm$ 1.3) $\times$ 10$^{18}$ \\
III    & 596     $\pm$ 1 & $120 \pm 5$ &  5   $\pm$  2   & 0.18 $\pm$ 0.02 & 1.2 $\pm$ 0.4 &  (2.2 $\pm$ 0.7) $\times$ 10$^{18}$ \\
Broad & 578 $\pm$ 4 & $ 142 \pm 5$ & 53   $\pm$ 10   & 0.01 $\pm$ 0.02 & 0.5 $\pm$ 1.1  &  (1   $\pm$ 2) $\times$ 10$^{18}$ \\
\tableline

d) \\
IV     & 540 $\pm$ 1 &  $546 \pm 10$ &  3   $\pm$  2   & 0.10 $\pm$ 0.03 & 0.5 $\pm$ 0.2 &  (0.9 $\pm$ 0.4) $\times$ 10$^{18}$ \\ 
SV    & 551 $\pm$ 1 & $286 \pm 10$ &  7.7 $\pm$  1.3 & 0.75 $\pm$ 0.03 & 6.7 $\pm$ 1.1 & (12   $\pm$ 2) $\times$ 10$^{18}$ \\
I/II   & 574 $\pm$ 1 & $400 \pm 10$ & 12.8 $\pm$  1.8 & 0.41 $\pm$ 0.03 & 5.9 $\pm$ 0.9 & (10.7 $\pm$ 1.7) $\times$ 10$^{18}$ \\
III    & 595 $\pm$ 1 & $506 \pm 10$ &  5.6 $\pm$  1.6 & 0.18 $\pm$ 0.03 & 1.5 $\pm$ 0.4 &  (2.7 $\pm$ 0.7) $\times$ 10$^{18}$ \\
Broad & 577 $\pm$ 4 &  $564 \pm 10$ & 55   $\pm$ 10   & 0.06 $\pm$ 0.03 & 3.6 $\pm$ 1.3 &  (7  $\pm$ 2) $\times$ 10$^{18}$ \\

\tableline
\end{tabular}
\tablenotetext{a}{ Velocity ($V$), flux density peak (Peak $S$),  FWHM (full width half maximum), peak optical depth (Peak $\tau$) and integrated optical depth ($\int \tau$ d$V$) were derived from Gaussian fits. Velocities in parenthesis are assumed values in the fitting. Uncertainties in the flux density peak and the peak optical depth correspond to the residuals after Gaussian fitting (see Fig.~ \ref{fig6-6}). Uncertainties in the integrated optical depth are calculated as the uncertainty of the area of the Gaussian profile. Upper limits in the integrated optical depth are $3\sigma$, and are calculated as 3 $\times$ $\sigma$ $\times$ ($\Delta V$ $\delta v$)$^{1/2}$, where $\Delta V$ is the velocity of the integrated profile and $\delta v$ the channel spacing. 
Line SV in position $a$ is detected, but the optical depth cannot be calculated since the absorption depth is larger than the continuum level due to the noise. Thus, we consider it as an upper limit.}
\end{center}
\end{table}

\begin{table}
\begin{center}

\caption{Parameters of the CO(2--1), $^{13}$CO(2-1) and C$^{18}$O(2-1) absorption lines  \label{tbl-4}}
\begin{tabular}{cccccccc}
\tableline
\tableline

Isotope           & Line & $V$          & $S$ & Peak $\tau$ &$\int \tau$dV   & $N(\rm CO)$$^{(a)}$ & $N(\rm H_2)$ $^{(b)}$ \\
                &                   & (km s$^{-1}$) &    (Jy beam$^{-1}$)     &        & (km s$^{-1}$) & (10$^{14}$ cm$^{-2}$) & (10$^{18}$ cm$^{-2}$)\\
\tableline                                                                        
\\
$^{12}$CO(2-1)              & \emph{1}        & 539           & 0.72  $\pm$ 0.25 & 2.1   $\pm$ 0.3 & 6.00  $\pm$ 0.10  &    616 $\pm$ 10         &     725 $\pm$ 12          \\
                           & \emph{2}         & 544          & 1.20  $\pm$ 0.25 & 1.6   $\pm$ 0.2 & 2.33  $\pm$ 0.07  &     239  $\pm$ 7        &    281  $\pm$ 8         \\
                           & \emph{3}         & 550          &  0.16  $\pm$ 0.25 & $\geq$ 3.6     & $\geq$ 14.88      &   $\geq$  1530          &    $\geq$ 1800         \\
                           & \emph{4}         & 575          & 5.33  $\pm$ 0.25 & 0.10 $\pm$ 0.05 & 0.22  $\pm$ 0.03  &      23 $\pm$ 3         &     26 $\pm$ 4         \\
                           & \emph{5}         & 606          & 2.66  $\pm$ 0.25 & 0.80 $\pm$ 0.10 & 1.68  $\pm$ 0.07  &    172  $\pm$ 7         &    203  $\pm$ 8       \\
                           & \emph{6}         & 613          & 4.61  $\pm$ 0.25 & 0.25 $\pm$ 0.06 & 0.48  $\pm$ 0.07  &    49  $\pm$ 7          &   58  $\pm$ 8        \\
\tableline                                                                                                                                      
\\                                                                                                                                              
$^{13}$CO(2-1)             & \emph{1}          & 541          & 5.30 $\pm$ 0.10 &0.11 $\pm$ 0.02   & 0.35 $\pm$ 0.03   &  36  $\pm$ 3       &   3600  $\pm$ 300    \\
                           & \emph{2}          & 545          & 4.90 $\pm$ 0.10 &0.19  $\pm$ 0.02 & 0.46 $\pm$ 0.03   &    47 $\pm$ 3      &   4700 $\pm$ 300          \\
                           & \emph{3}          & 552          &3.33  $\pm$ 0.10 &0.58  $\pm$ 0.03 & 1.42 $\pm$ 0.04   &    145 $\pm$ 4     &   14500  $\pm$ 400     \\
                           & \emph{4}          & --           &  --             & --              & $<0.08$           &       $<8$         &   $<800$               \\
                           & \emph{5}          & 607          & 5.80  $\pm$ 0.10& 0.05 $\pm$ 0.01 & 0.10 $\pm$ 0.02   &     10  $\pm$ 2    &   1000    $\pm$ 200      \\
                           & \emph{6}          & --           & --              &  --             & $<0.09$           &     $<9$           &   $<900$             \\
\tableline                                                                                                                                      
\\                                                                                                                                              
C$^{18}$O(2-1)             & --                 & --          &   --            &  --              &$<0.11$           &  $<$ 11   &  $<6600$             \\
\tableline

\end{tabular}
\tablenotetext{a}{Isotopic CO column densities.We assume a filling factor of unity, local thermodynamic equilibrium (LTE) and an excitation temperature $T_{\rm ex}$ = 10~K  for the absorbing molecular gas (\S~\ref{sub:optdepthColden}). Note that the CO(2--1) line \emph{3} is likely saturated, so the optical depth and the column densities are lower limits. 
 $^{13}$CO lines \emph{4} and \emph{6} and C$^{18}$O lines are not detected, thus we use upper limits  taken at a 3~$\sigma$ level.}
\tablenotetext{b}{H$_2$ column density ($N(\rm H_2)$)  derived from the CO(2--1) absorption lines (and isotopes). To derive the $N(\rm H_2)$ from the CO(2--1) lines, we use an abundance ratio of [CO]/[H$_2$] = 8.5 $\times$ 10$^{-5}$ \citep{1982ApJ...262..590F}, [$^{13}$CO]/[H$_2$] = 1 $\times$ 10$^{-6}$ \citep{1979ApJ...232L..89S} and [C$^{18}$O]/[H$_2$] = 1.7 $\times$ 10$^{-7}$ \citep{1982ApJ...262..590F} ([CO]/[$^{13}$CO] = 85 and [CO]/[C$^{18}$O] = 500]).}
\end{center}
\end{table}

\normalsize

\end{document}